\documentclass[12pt,thmsa]{article}
\usepackage{amsfonts}

\input{tcilatex}
 \usepackage[T1]{fontenc}
\begin{document}

\date{Sept. 21, 2005 }
\title{\textbf{Twisting Null Geodesic Congruences and the Einstein-Maxwell Equations%
}}
\author{Ezra T.\ Newman$^{1}$ and Gilberto Silva-Ortigoza$^{2}$ \\
$^{1}$Dept of Physics and Astronomy, \\
Univ. of Pittsburgh, \\
Pittsburgh, PA 15260, USA\\
$^{2}$Facultad de Ciencias F\'{\i }sico Matem\'{a}ticas \\
de la Universidad Aut\'{o}noma de Puebla, \\
Apartado Postal 1152, \\
Puebla, Pue., M\'{e}xico}
\maketitle

\begin{abstract}
In a recent article we returned to the study of asymptotically flat
solutions of the vacuum Einstein equations with a rather unconventional
point of view. The essential observation in that work was that from a given
asymptotically flat vacuum space-time with a given Bondi shear, one can find
a class of asymptotically shear-free (but, in general, twisting) null
geodesic congruences where the class was uniquely given \textit{up to the
arbitrary choice of a complex analytic `world-line' in a four- dimensional
complex space}. By imitating certain terms in the Weyl tensor that are found
in the algebraically special type II metrics, this complex world-line could
be made unique and given - or assigned - the physical meaning as the complex
center of mass. Equations of motion for this case were found.

The purpose of the present work is to extend those results to asymptotically
flat solutions of the Einstein-Maxwell equations. Once again, in this case,
we get a class of asymptotically shear-free null geodesic congruences
depending on a complex world-line in the same four-dimensional complex
space. However in this case there will be, in general, two distinct but
uniquely chosen world-lines. One of which can be assigned as the complex
center-of- charge while the other could be called the complex center of
mass. Rather than investigating the situation where there are two distinct
complex world-lines, we study instead the special degenerate case where the
two world-lines coincide, i.e., where there is a single unique world-line.
This mimics the case of algebraically special Einstein-Maxwell fields where
the degenerate principle null vector of the Weyl tensor coincides with a
Maxwell principle null vector. Again we obtain equations of motion for this
world-line - but explicitly found here only in an approximation. Though
there are ambiguities in assigning physical meaning to different terms it
appears as if reliance on the Kerr and charged Kerr metrics and classical
electromagnetic radiation theory helps considerably in this identification.
In addition, the resulting equations of motion appear to have many of the
properties of a particle with intrinsic spin and an intrinsic magnetic
dipole moment. At first order there is even the classical radiation-reaction
term $\frac{2}{3}\frac{q^{2}}{c^{3}}\ddot{v},$ now obtained without any use
of the Lorentz force law but obtained directly from the asymptotic fields
themselves. One even sees the possible suppression of the classical runaway
solutions due to the radiation reaction force.
\end{abstract}

\section{\textbf{Introduction}}

In a recent article\cite{CTG3} we returned, with a rather unconventional
point of view, to the study of asymptotically flat solutions of the vacuum
Einstein equations . The main development in that work was the realization
that for \textit{any} given asymptotically flat space-time with a given
Bondi asymptotic shear, one can find a class of asymptotically shear-free
(but, in general, twisting) null geodesic congruences where the class was
uniquely given \textit{up to an arbitrary choice of a complex analytic
`world-line' in a four- complex dimensional space}. Furthermore, by
mimicking some terms that are found in the Weyl tensor of the algebraically
special type II metrics, this complex world-line could be chosen uniquely
and given - or assigned - the physical meaning of a complex center of mass;
the real part being related to an asymptotically defined center of mass and
the imaginary part related to asymptotically defined intrinsic spin.
Equations of motion for the real and imaginary parts were found to strongly
resemble the Mathisson-Papapetrou equations of motion for a spinning
particle.

The purpose of the present work is to extend those results to asymptotically
flat solutions of the Einstein-Maxwell equations. Once again, in this case,
we get a class of asymptotically shear-free null geodesic congruences
depending on a complex world-line in the same four-dimensional complex
space. However in this case there will be, in general, two distinct but
uniquely given world-lines. One, which comes from properties of the Maxwell
field, can be assigned as the complex center-of-charge while the other
arising from properties of the Weyl tensor, can be associated with the
complex center of mass. Rather than study the situation where there are two
distinct complex world-lines, we study instead the special degenerate case
where the two world-lines coincide, i.e., were there is a single unique
world-line. This mimics the case of algebraically special Einstein-Maxwell
fields where the degenerate principle null vector of the Weyl tensor
coincides with a Maxwell principle null vector. Again we obtain equations of
motion for this world-line - found here only in a second-order
approximation. Though there is not a normal or standard method for the
assignment of physical meaning to different terms or quantities,
nevertheless tentative assignments can be made by a variety of methods. One
can use the Kerr and charged Kerr metrics to identify spin and magnetic
moments. From the equations of motion that appear to have many of the
properties of a particle with intrinsic spin and an intrinsic magnetic
dipole moment \{the Mathisson-Papapetrou equations\}, one can try to
identify asymptotically defined centers of mass and charge. One even has at
first order, the classical radiation-reaction term $\frac{2}{3}\frac{q^{2}}{%
c^{3}}\ddot{v}$ now obtained without any use of the Lorentz force law but
simply from the asymptotic fields themselves. A new development occurs when
we go to a 3rd order term in the equations of motion: we find a counter term
that appears to suppress or dampen the exponential run-away solutions
associated with the radiation reaction force. This arises solely from the
Bondi mass loss equation.

One should think of this work as developing a generalization of the
properties of the algebraically special space-times in the sense that the
term that is required here to vanish, is automatically vanishing (among many
other terms) for all the algebraically special metrics. It was, in fact, an
understanding of the algebraically special metrics and their associated 
\textit{shear-free null congruence }that led us to this construction of the 
\textit{asymptotically shear-free} congruences and the unique complex
world-line.

The charged Robinson-Trautman metrics and the charged Kerr metrics with
their properties are explicit examples of the ideas and construction given
here.

In Section II, which is a summary of our earlier review\cite{CTG3}, we
discuss certain properties of null infinity $\frak{I}^{+}$ that are needed
here. In Section III and IV we describe the variables that come from the
Maxwell and Weyl tensors and live on $\frak{I}^{+}$ and their evolution,
i.e., the asymptotic Maxwell equations and Bianchi Identities$.$ In Section
V we describe how to find the family of asymptotically shear free
congruences, i.e., how to determine a field of complex stereographic angles, 
$L[\frak{I}^{+}],$ and then how to narrow that family down to two specific
congruences: one comes from the Maxwell tensor while the other comes from
the Weyl tensor. In the present work we shall assume that we are dealing
with the special case when these two world-line coincide. We thus find, in
this case, that there is a unique complex world-line in H-space that
determines the unique shear-free null geodesic congruence. Finally, in
Section VI, we attempt to give physical meaning and significance to the
geometric structures that have arisen in this work.

In particular, as a great surprise to us, we find that the equations of
motion for this complex world-line are extremely close to the standard
equations of motion for a charged particle with intrinsic spin. The real
part appears to represent a description of the motion of the center of mass
of the interior gravitating system and the imaginary part the evolution of
the spin vector of the same system.

There appears to us a genuine mystery here: Why should the motion in a
complex parameter space be virtually the same as ordinary motion in
space-time?

It is unfortunate, but apparently unavoidable, that there is considerable
duplication of the material presented here with that of our earlier work\cite
{CTG3}. Not to have had this duplication would, we believe, make the present
work unintelligible.

\section{\textbf{Review of }$\frak{I}^{+}$\textbf{: the Future Null Boundary
of Space-Time}}

\subsection{$\frak{I}^{+}$ coordinates}

Future null infinity, often referred to as $\frak{I}^{+},$ is roughly
speaking the set of endpoints of all future directed null geodesics\cite{P2}%
. It is usually given the structure of $S^{2}xR,$ a line bundle over the
sphere. This leads naturally to the global stereographic coordinates, ($%
\zeta ,\overline{\zeta }),$ for the $S^{2}$ and with $u$ labeling the cross
sections or global slices of the bundle. From the point of view of the
space-time, $\frak{I}^{+}$ is a null surface, with the null generators given
by ($\zeta ,\overline{\zeta })$ $=constant$.

There is a canonical slicing of $\frak{I}^{+}$, $u_{B}=constant$, referred
to as the Bondi slicing, that it is defined from an asymptotic symmetry
inherited from the interior, called the Bondi-Metzner-Sachs group\cite{P3}.
Any one Bondi slicing is related to any other by the \textit{supertranslation%
} freedom;

\begin{equation}
\widetilde{u}_{B}=u_{B}+\alpha (\zeta ,\overline{\zeta }).  \label{ST}
\end{equation}
with $\alpha (\zeta $,$\overline{\zeta })$ an arbitrary regular function on $%
S^{2}.$

\quad All other arbitrary slicings, ($u=\tau ),$ often called NU slicings,
are given by 
\begin{equation}
\quad u_{B}=X(\tau ,\zeta ,\overline{\zeta }).  \label{NUT}
\end{equation}

\subsection{Null Tetrads on $\frak{I}^{+}$}

In addition to the choice of coordinate systems there is the freedom to
chose the asymptotic null tetrad system, ($l^{a},m^{a},\overline{m}%
^{a},n^{a} $). Since $\frak{I}^{+}$ is a fixed null surface with its own
generators, with null tangent vectors, [say $n^{a}],$ it is natural to keep
it fixed. The remaining tetrad freedom lies, essentially, in the choice of
the other null leg, $l^{a}.$ Most often $l^{a}$ is chosen to be orthogonal
to the Bondi, $u_{B}=constant,$ slices. In this case we will refer to a
Bondi coordinate/tetrad system. Relative to such a system any other tetrad
set, ($l^{*a},m^{*a},n^{*a}$), is given by a null rotation\cite{Aronson}
about $n^{a},$ i.e., 
\begin{eqnarray}
l^{*a} &=&l^{a}+b\overline{m}^{a}+\overline{b}m^{a}+b\overline{b}n^{a},
\label{NullRot} \\
m^{*a} &=&m^{a}+bn^{a},  \nonumber \\
n^{*a} &=&n^{a},  \nonumber \\
b &=&-L/r+O(r^{-2}).  \nonumber
\end{eqnarray}

The \textit{complex function} $L(u_{B},\zeta ,\overline{\zeta })$, the
(complex stereographic) angle between the null vectors, $l^{*a}$ and $l^{a}$
at each point on $\frak{I}^{+}$, is at this moment a completely arbitrary
angle field but later will be dynamically determined. Depending on how $%
L(u_{B},\zeta ,\overline{\zeta })$ is chosen $l^{*a}$ might or might not be
surface forming. We will abuse the language/notation and refer to the tetrad
systems associated with $l^{*a}$ as the \textit{twisting-type} tetrads as
distinct from a Bondi tetrad even when $l^{*a}$ is surface forming.

The function $L(u_{B},\zeta ,\overline{\zeta })$ and its choice will later 
\textit{play the pivotal role} in this work. It will be seen that the vector
field $l^{*a}$ can be constructed by an appropriately chosen $L(u_{B},\zeta ,%
\overline{\zeta })$ so that it is asymptotically \textit{shear-free}.

\section{\textbf{Quantities Defined on }$\frak{I}^{+}$\textbf{\ for Any
Given Interior Einstein-Maxwell Space-Time}}

Using the spin-coefficient notation for the three complex components of the
asymptotic Maxwell field and for the asymptotic Weyl tensor, we have, from
the peeling theorem\cite{TN,NP}, that:

\begin{eqnarray}
\phi _{0} &=&\frac{\phi _{0}^{0}}{r^{3}}+0(r^{-4}) \\
\phi _{1} &=&\frac{\phi _{1}^{0}}{r^{2}}+0(r^{-3}) \\
\phi _{2} &=&\frac{\phi _{2}^{0}}{r}+0(r^{-2})
\end{eqnarray}

\begin{eqnarray}
\psi _{0} &=&\frac{\psi _{0}^{0}}{r^{5}}+0(r^{-6}) \\
\psi _{1} &=&\frac{\psi _{1}^{0}}{r^{4}}+0(r^{-5}) \\
\psi _{2} &=&\frac{\psi _{2}^{0}}{r^{3}}+0(r^{-4}) \\
\psi _{3} &=&\frac{\psi _{3}^{0}}{r^{2}}+0(r^{-3}) \\
\psi _{4} &=&\frac{\psi _{4}^{0}}{r}+0(r^{-2})
\end{eqnarray}
where $\phi _{0}^{0},\phi _{1}^{0},\phi _{2}^{0},\psi _{0}^{0},\psi
_{1}^{0},\psi _{2}^{0},\psi _{3}^{0},$ and $\psi _{4}^{0}$ are functions
defined on $\frak{I}^{+}.$ For a given space-time their explicit expressions
depend on the choices of \textit{both} coordinates and tetrads. Their
evolution is determined by the asymptotic Einstein-Maxwell equations and the 
\textit{choice of characteristic data}.\newline

We adopt the following notation: expression written in the Bondi
coordinate/tetrad system will appear without a star,`$^{*}$', e.g., $\psi
_{2}^{0},$ while for Bondi coordinates with a twisting tetrad a `$^{*}$'
will be used, e.g., $\psi _{2}^{*0}$. In the case of NU coordinates and
twisted tetrad, a double star will be used, e.g. $\psi _{2}^{**0},$etc.

Using Bondi coordinates and tetrad, the free characteristic data is given
only by the (complex) asymptotic shear,

\[
\sigma =\sigma (u_{B},\zeta ,\overline{\zeta }), 
\]
while in Bondi coordinates, but with a twisting-type tetrad the free
functions are 
\[
\sigma ^{*}=\sigma ^{*}(u_{B},\zeta ,\overline{\zeta })\,\text{ \& }
\,L(u_{B},\zeta ,\overline{\zeta }) 
\]
with $\sigma ^{*}(u_{B},\zeta ,\overline{\zeta })$ and $L(u_{B},\zeta ,%
\overline{\zeta })$ carrying the same (redundant) information as did the
Bondi $\sigma (u_{B},\zeta ,\overline{\zeta }).$

In the case of NU coordinates and twisted tetrad, the free data is given by

\[
\sigma ^{**}=\sigma ^{**}(\tau ,\zeta ,\overline{\zeta })\text{\ \ \ \&\ }
\,V(\tau ,\zeta ,\overline{\zeta }) 
\]
with 
\[
V(\tau ,\zeta ,\overline{\zeta })\equiv du_{B}/d\tau =\partial _{\tau
}G(\tau ,\zeta ,\overline{\zeta }). 
\]

The important point is that in both cases, the Bondi coordinates with
twisting tetrad and the NU coordinates and twisted tetrad, all the
information that was in $\sigma (u_{B},\zeta ,\overline{\zeta })$ is
transferred to the new variables, ($\sigma ^{*}$\& $L)$ or ($\sigma ^{**}$\& 
$V)$. Later we will show that \textit{all the information} can be shifted
into an appropriately chosen $L(u_{B},\zeta ,\overline{\zeta }),$ \{which is
also a complex function on $\frak{I}^{\frak{+}}\frak{\}}$ with a vanishing $%
\sigma ^{*}$ and that in certain special cases (a pure `electric' type $%
\sigma $) \textit{all the information} can be shifted into the $V(\tau
,\zeta ,\overline{\zeta })$ with vanishing $\sigma ^{**}.$

The relationship between the ($\phi^0_2$, $\phi^0_1$ $\phi^0_0$), ($\psi
_{4}^{0},\psi _{3}^{0},\psi _{2}^{0},\psi _{1}^{0},\psi _{0}^{0})$ given in
a Bondi tetrad and the ($\phi^{0*}_2$, $\phi^{0*}_1$, $\phi^{0*}_0$), ($\psi
_{4}^{*0},\psi _{3}^{*0},\psi _{2}^{*0},\psi _{1}^{*0},\psi _{0}^{*0})$ of a
twisting tetrad is

\begin{eqnarray}
\phi^0_0 & = & \phi^{*0}_0 + 2 L \phi^{*0}_1 + L^2 \phi^{*0}_2,
\label{BtoTphi00} \\
\phi^0_1 & = & \phi^{*0}_1 + L \phi^{*0}_2,  \label{BtoTphi01} \\
\phi^{0}_2 & = & \phi^{*0}_2,  \label{BtoTphi02}
\end{eqnarray}
\begin{eqnarray}
\psi _{0}^{0} &=&\psi _{0}^{*0}+4L\psi _{1}^{*0}+6L^{2}\psi
_{2}^{*0}+4L^{3}\psi _{3}^{*0}+L^{4}\psi _{4}^{*0},  \label{BtoTpsi00} \\
\psi _{1}^{0} &=&\psi _{1}^{*0}+3L\psi _{2}^{*0}+3L^{2}\psi
_{3}^{*0}+L^{3}\psi _{4}^{*0},  \label{BtoTpsi01} \\
\psi _{2}^{0} &=&\psi _{2}^{*0}+2L\psi _{3}^{*0}+L^{2}\psi _{4}^{*0},
\label{BtoTpsi02} \\
\psi _{3}^{0} &=&\psi _{3}^{*0}+L\psi _{4}^{*0},  \label{BtoTpsi03} \\
\psi _{4}^{0} &=&\psi _{4}^{*0}.  \label{BtoTpsi04}
\end{eqnarray}

These are used, in the next section, going from the Bondi version of the
Bianchi identities to the twisting version.

\section{The Asymptotic Maxwell equations and the Bianchi Identities}

The Asymptotic Maxwell equations and the Bianchi Identities in a Bondi
coordinate and tetrad system are:\newline

A. In a Bondi coordinate and tetrad system\cite{TN}:

\qquad \underline{Maxwell Equations} 
\begin{eqnarray}
&&\phi _{0}^{0\cdot }+\text{\dh }\phi _{1}^{0}-\sigma \phi _{2}^{0}=0,
\label{phiB00cdot} \\
&&\phi _{1}^{0\cdot }+\text{\dh }\phi _{2}^{0}=0.  \label{phiB10cdot}
\end{eqnarray}

\qquad \underline{Bianchi Identities}

\begin{eqnarray}
\psi _{0}^{0\,\cdot } &=&-\text{\dh }\psi _{1}^{0}+3\sigma \psi
_{2}^{0}+3k\phi _{0}^{0}\overline{\phi }_{2}^{0}  \label{Bondi1} \\
\psi _{1}^{0\,\cdot } &=&-\text{\dh }\psi _{2}^{0}+2\sigma \psi
_{3}^{0}+2k\phi _{1}^{0}\overline{\phi }_{2}^{0}  \label{Bondi2} \\
\psi _{2}^{0\,\cdot } &=&-\text{\dh }\psi _{3}^{0}+\sigma \psi
_{4}^{0}+k\phi _{2}^{0}\overline{\phi }_{2}^{0}  \label{Bondi3} \\
\psi _{3}^{0} &=&\text{\dh }\overline{\sigma }^{\cdot }  \label{Bondi4} \\
\psi _{4}^{0} &=&-\overline{\sigma }^{\cdot \cdot }  \label{Bondi5} \\
\psi _{2}^{0\,}-\overline{\psi }_{2}^{0\,} &=&\overline{\text{\dh }}%
^{2}\sigma -\text{\dh }^{2}\overline{\sigma }+\overline{\sigma }\sigma
^{\cdot }-\sigma \overline{\sigma }^{\cdot },  \label{Bondi6}
\end{eqnarray}
where 
\begin{equation}
k=\frac{2G}{c^{4}}.  \label{k}
\end{equation}

By introducing the definition of the mass aspect 
\begin{equation}
\Psi =\psi _{2}^{0\,}+\text{\dh }^{2}\overline{\sigma }+\sigma \overline{
\sigma }^{\cdot },  \label{Psi}
\end{equation}
we see that (\ref{Bondi6}) is equivalent to 
\begin{equation}
\overline{\Psi }=\Psi .  \label{PsiReal}
\end{equation}

In the following section we show how the function $L(u_{B},\zeta ,\overline{
\zeta })$ can be chosen so that the shear $\sigma ^{*},$ of the twisting
congruence vanishes. For the moment we just assume that $\sigma ^{*}=0$ for
the twisting tetrad.

By using Eqs.\thinspace (\ref{BtoTphi00})-(\ref{BtoTpsi04}) we find that for
a Bondi Coordinate/Twisting Tetrad system with $\sigma ^{*}=0,$ the same set
of equations, the Maxwell and Bianchi identities, can be rewritten as:%
\newline

B. In Bondi coordinates with a twisting tetrad and $\sigma ^{*}=0:$

\qquad \underline{Maxwell Equations} 
\begin{eqnarray}
\phi _{0}^{*0\cdot }+\text{\dh }\phi _{1}^{*0}+2L^{\cdot }\phi
_{1}^{*0}+L\phi _{1}^{*0\cdot } &=&0,  \label{ME1T} \\
\phi _{1}^{*0\cdot }+\text{\dh }\phi _{2}^{*0}+(L\phi _{2}^{*0})^{\cdot }
&=&0.  \label{ME2T}
\end{eqnarray}

\qquad \underline{Bianchi Identities}

\begin{eqnarray}
\psi _{0}^{*0\,\cdot } &=&-\text{\dh }\psi _{1}^{*0}-L\psi _{1}^{*0\,\cdot
}-4L^{\cdot }\psi _{1}^{*0}+3k\phi _{0}^{*0}\overline{\phi }_{2}^{*0}
\label{twisting1} \\
\psi _{1}^{*0\,\cdot } &=&-\text{\dh }\psi _{2}^{*0}-L\psi _{2}^{*0\,\cdot
}-3L^{\cdot }\psi _{2}^{*0}+2k\phi _{1}^{*0}\overline{\phi }_{2}^{*0}
\label{twisting2} \\
\psi _{2}^{*0\,\cdot } &=&-\text{\dh }\psi _{3}^{*0}-L\,\psi _{3}^{*0\,\cdot
}-2L^{\cdot }\psi _{3}^{*0}+k\phi _{2}^{*0}\overline{\phi }_{2}^{*0}
\label{twisting3} \\
\psi _{3}^{*0} &=&\text{\dh }\overline{\sigma }^{\cdot }+L\overline{\sigma }%
^{\cdot \cdot }  \label{twisting4} \\
\psi _{4}^{*0} &=&-\overline{\sigma }^{\cdot \cdot }  \label{twisting5} \\
\psi _{2}^{*0}-\overline{\psi }_{2}^{*0} &=&\overline{\text{\dh }}^{2}\sigma
-\text{\dh }^{2}\overline{\sigma }+\overline{\sigma }\sigma ^{\cdot }-\sigma 
\overline{\sigma }^{\cdot }  \nonumber \\
&&+2\overline{L}\,\overline{\text{\dh }}\sigma ^{\cdot }-2L\text{\dh }%
\overline{\sigma }^{\cdot }+\overline{L}^{2}\sigma ^{\cdot \cdot }-L^{2}%
\overline{\sigma }^{\cdot \cdot },  \label{Bondi6*}
\end{eqnarray}
where 
\begin{equation}
\sigma \equiv \text{\dh }L+LL^{\cdot }.  \label{twisting6}
\end{equation}
By using (\ref{BtoTpsi02}) in (\ref{Psi}) we finds that 
\begin{equation}
\Psi =\psi _{2}^{*0}+2L\text{\dh }\overline{\sigma }^{\cdot }+L^{2}\overline{%
\sigma }^{\cdot \cdot }+\text{\dh }^{2}\overline{\sigma }+\sigma \overline{
\sigma }^{\cdot }.  \label{Psi*}
\end{equation}
so that (\ref{Bondi6*}) is again equivalent to 
\begin{equation}
\Psi -\overline{\Psi }=0.  \label{twisting8}
\end{equation}
The dot denotes derivative with respect to $u_{B}$.\newline

In the very special case where the Bondi shear is of pure `electric type',
the `twist' of the asymptotically shear-free null congruence $l^{*a}$ can be
chosen to vanish and hence the $l^{*a}$ would be orthogonal to the cuts of
some constant $\tau $ slices of $\frak{I}^{+}$ . The slices, given by $\tau
=T(u_{B},\zeta ,\overline{\zeta })$ or $u_{B}=X(\tau ,\zeta ,\overline{\zeta 
})$, yield the parametric relations, $($see below, Eqs.(\ref{tau}), (\ref{X}
) and (\ref{L2})) 
\begin{eqnarray*}
L(u_{B},\zeta ,\overline{\zeta }) &=&\text{\dh }_{(\tau )}X(\tau ,\zeta ,%
\overline{\zeta }) \\
V(u_{B},\zeta ,\overline{\zeta }) &=&\text{\dh }_{\tau }X(\tau ,\zeta ,%
\overline{\zeta })\equiv X^{\prime } \\
u_{B} &=&X(\tau ,\zeta ,\overline{\zeta }).
\end{eqnarray*}

To perform the coordinate transformation to ($\tau $, $\zeta $, $\overline{%
\zeta }$) \{from ($u_{B}$, $\zeta $, $\overline{\zeta }$)\}, referred to as
NU coordinates, of the Maxwell equations and Bianchi identities, it is
convenient to note that 
\begin{eqnarray}
\text{\dh }\eta &=&\text{\dh }_{(\tau )}\eta -\frac{L}{V}\eta ^{\prime } 
\nonumber \\
\eta ^{\cdot } &=&\frac{1}{V}\eta ^{\prime }
\end{eqnarray}
where \dh $_{(\tau )}$ is the edth operator taking $\tau $ constant, $\eta $
is a function with spin weight $s$ and $\eta ^{\prime }$ is the derivative
of $\eta $ with respect to $\tau $. Then we have:\newline

C. In NU coordinates and `twisting' tetrad with $\sigma ^{**}=0$:

\qquad \underline{Maxwell Equations} 
\begin{eqnarray}
V\phi _{0}^{**0\prime }+\text{\dh }_{(\tau )}[V^{2}\phi _{1}^{**0}] &=&0, \\
\phi _{1}^{**0\prime }+\text{\dh }_{(\tau )}[V\phi _{2}^{**0}] &=&0.
\end{eqnarray}

\qquad \underline{Bianchi Identities}

\begin{eqnarray}
\psi _{0}^{**0\,\prime } &=&-V\text{\dh }_{(\tau )}\psi _{1}^{**0}-4[\text{
\dh }_{(\tau )}V]\psi _{1}^{**0}+3kV\phi _{0}^{**0}\overline{\phi }_{2}^{**0}
\\
\psi _{1}^{**0\,\prime } &=&-V\text{\dh }_{(\tau )}\psi _{2}^{**0}-3[\text{%
\dh }_{(\tau )}V]\psi _{2}^{**0}+2kV\phi _{1}^{**0}\overline{\phi }_{2}^{**0}
\nonumber \\
\psi _{2}^{**0\,\prime } &=&-V\text{\dh }_{(\tau )}\psi _{3}^{**0}-2[\text{
\dh }_{(\tau )}V]\psi _{3}^{**0}+kV\phi _{2}^{**0}\overline{\phi }_{2}^{**0}
\nonumber \\
\psi _{3}^{**0} &=&V^{-1}[\overline{\text{\dh }}_{(\tau )}^{2}\text{\dh }
_{(\tau )}V+2\overline{\text{\dh }}_{(\tau )}V]-V^{-2}[\overline{\text{\dh }}
_{(\tau )}^{2}V][\text{\dh }_{(\tau )}V]  \nonumber \\
\psi _{4}^{**0} &=&V^{-3}[V^{\prime }\overline{\text{\dh }}_{(\tau )}^{2}V-V%
\overline{\text{\dh }}_{(\tau )}^{2}V^{\prime }]  \nonumber \\
\psi _{2}^{**0}-\overline{\psi }_{2}^{**0} &=&0.  \nonumber
\end{eqnarray}

We emphasize that we have not lost any generality when going to the
Bondi/Twisting tetrad but when we change to the NU coordinates the equations 
\textit{are valid only when the original Bondi shear} $\sigma (u_{B},\zeta ,%
\overline{\zeta })$ \textit{was pure `electric'}. In other words all the
information in an `electric' shear can be put into the real $V(u_{B},\zeta ,%
\overline{\zeta }$). Though in the present work these equations in the NU
coordinates will not be used, they are included here for future use with
further applications, including an analysis of the (charged)
Robinson-Trautman\cite{RT} equations.

\section{\textbf{Dynamics on }$\frak{I}^{+}$}

\subsection{The asymptotic shear-free condition}

We first show that there are special choices of $L(u_{B},\zeta ,\overline{
\zeta })$ leading to the twisting tetrads for which the shear, $\sigma
^{*}=0.$\ To find this family of functions $L(u_{B},\zeta ,\overline{\zeta }%
),$ a differential equation must be solved where the freedom in the solution
is four arbitrary complex analytic functions of a single complex variable,
i.e., an arbitrary complex analytic curve in a four-complex dimensional
parameter space. [Surprisingly, the parameter space is the well-studied $H$
-space.] All the information in the original Bondi characteristic data, $%
\sigma (u_{B},\zeta ,\overline{\zeta }),$ will have been shifted to the $%
L(u_{B},\zeta ,\overline{\zeta }).$ Given an $L(u_{B},\zeta ,\overline{\zeta 
})$ with \textit{any one of these curves,} we could go backwards to recover
the original Bondi $\sigma (u_{B},\zeta ,\overline{\zeta }).$ The dynamics
lies in the unique determination of this curve from other considerations
that are describe later. The construction of these quantities, the $%
L(u_{B},\zeta ,\overline{\zeta }),$ the complex curve, the $l^{*},$ etc.,
though done in a particular Bondi coordinate system, are invariant under
supertranslations.

We begin by observing that if we start with a Bondi coordinate/tetrad system
with a given shear $\sigma (u_{B},\zeta ,\overline{\zeta }),$ then [after an
unpleasant calculation\cite{Aronson}] the shear of the new null vector $%
l^{*a},$ after the null rotation, Eqs.(\ref{NullRot}), is related to the old
one by 
\[
\sigma ^{*}(u_{B},\zeta ,\overline{\zeta })=\sigma (u_{B},\zeta ,\overline{%
\zeta })-\text{\dh }L-LL^{\cdot }. 
\]

The function $L(u_{B},\zeta ,\overline{\zeta })$ is then chosen so that the
new shear vanishes, and hence, it must satisfy 
\begin{equation}
\text{\dh }L+LL^{\cdot }=\sigma (u_{B},\zeta ,\overline{\zeta }).
\label{ShearFree}
\end{equation}

\begin{remark}
The special case, \dh $L+LL^{\cdot }=0,$ played an important early role in
the development of twistor theory, leading immediately to the relationship
in flat space between shear-free null geodesic congruences and twistor
theory. It leads us to conjecture that Eq.(\ref{ShearFree}) might play a
role in some asymptotic form of twistor theory.\newline
\end{remark}

Though Eq.(\ref{ShearFree}) is non-linear with an arbitrary right-side and
appears quite formidable, considerable understanding of it can be found by
the following procedure:

Assume the existence of a \textit{complex} analytic function of ($%
u_{B},\zeta ,\overline{\zeta })$ \{where $\overline{\zeta }$ is allowed to
be freed up from the complex conjugate of $\zeta $ and where $u_{B}$ can
take complex values\}, 
\begin{equation}
\tau =T(u_{B},\zeta ,\overline{\zeta })  \label{tau}
\end{equation}
that is invertible in the sense that it can, in principle, be written as 
\begin{equation}
u_{B}=X(\tau ,\zeta ,\overline{\zeta }).  \label{X}
\end{equation}

\begin{remark}
We emphasize the obvious: since $u_{B}$ is allowed to be complex, then the
function $X(\tau ,\zeta ,\overline{\zeta })$ is also complex.
\end{remark}

Then writing 
\begin{equation}
L=-\frac{\text{\dh }T}{T^{\cdot }},  \label{L1}
\end{equation}
and using the implicit derivatives of Eq.(\ref{X}), 
\begin{eqnarray}
1 &=&X,_{\tau }T^{\cdot }  \label{Implicit} \\
0 &=&\text{\dh }_{(\tau )}X+X^{\prime }\text{\dh }T,  \nonumber
\end{eqnarray}
we obtain 
\begin{equation}
L=\text{\dh }_{(\tau )}X.  \label{L2}
\end{equation}
Again we mention that prime means the $\tau $ derivative and \dh $_{(\tau )}$
means \dh\ with $\tau $ held constant. Finally, with this implicit view of $%
X $ and $T$, Eq.(\ref{ShearFree}) becomes 
\begin{equation}
\text{\dh }_{(\tau )}^{2}X=\sigma (X,\zeta ,\overline{\zeta }).
\label{GoodCut}
\end{equation}

\begin{remark}
This equation had been derived earlier and was referred to as the `good cut
equation'\cite{H-space1,H-space2}\textit{.} In general it has a four complex
dimensional solution space that has been named $H$-space. In its original
form it had its origin in the search for \textit{complex null surfaces} that
were asymptotically shear-free. Here we are looking for \textit{real} 
\textit{null geodesic congruences} \textit{that are, in general, }\underline{%
\textit{\textbf{not} surface forming}}, i.e., they have twist, but are
asymptotically shear-free\cite{KN2}.\newline
\end{remark}

\begin{theorem}
Note that using the reparametrization 
\begin{equation}
\tau \Rightarrow \tau ^{*}=F(\tau )=T^{*}(u_{B},\zeta ,\overline{\zeta })
\label{repar}
\end{equation}
leaves Eq.(\ref{L1}) and the entire construction invariant. This fact will
be used later for certain simplifications.\newline
\end{theorem}

\begin{remark}
We mention that if the shear in Eq.(\ref{GoodCut}), is pure electric
[i.e.,\thinspace $\sigma (u_{B},\zeta ,\overline{\zeta })=$\dh $%
^{2}S(u_{B},\zeta ,\overline{\zeta })$ with $S(u_{B},\zeta ,\overline{\zeta }%
)$ a \textit{real function}] then the associated $H$-space is flat and has a
real four-dimensional subspace that can be identified with Minkowski space. 
\newline
\end{remark}

As just mentioned, the solutions to Eq.(\ref{GoodCut}) depend on four
complex parameters, say $z^{a},$ and can be summarized by

\begin{equation}
u_{B}=X(z^{a},\zeta ,\overline{\zeta }).  \label{H}
\end{equation}
However, since we are interested in solutions of the form, Eq.(\ref{X}),
i.e., 
\begin{equation}
u_{B}=X(\tau ,\zeta ,\overline{\zeta }),  \label{X*}
\end{equation}
all we must do is chose in, $H$-space, an arbitrary complex world-line, $%
z^{a}=\xi ^{a}(\tau ),$ and substitute it into Eq.(\ref{H}) to obtain the
form (\ref{X*}). Furthermore, it can be seen that the solution, (\ref{X*}),
can be written \{with a special choice of H-space coordinates\}as 
\[
u_{B}=X(\tau ,\zeta ,\overline{\zeta })=\xi ^{a}(\tau )l_{a}(\zeta ,%
\overline{\zeta })+X_{l\geq 2}(\tau ,\zeta ,\overline{\zeta }) 
\]
with $X_{l\geq 2}$ containing spherical harmonics, $l\geq 2$ and 
\begin{equation}
l_{a}(\zeta ,\overline{\zeta })=\frac{\sqrt{2}}{2}(1,-\frac{\zeta +\overline{%
\zeta }}{1+\zeta \overline{\zeta }},i\frac{\zeta -\overline{\zeta }}{1+\zeta 
\overline{\zeta }},\frac{1-\zeta \overline{\zeta }}{1+\zeta \overline{\zeta }%
})  \label{X1.2}
\end{equation}
which has only the $l=0$ and $1,$ spherical harmonics. This result follows
from the fact that \dh $_{(\tau )}^{2}$ in the good-cut equation annihilates
the $l=0$ and $1$ terms. The $\xi ^{a}(\tau )$ does get fed back into the
higher harmonics via the $\sigma (X,\zeta ,\overline{\zeta }).$\newline

Our solution to Eq.(\ref{ShearFree}) can now be expressed implicitly by 
\begin{eqnarray}
L(u_{B},\zeta ,\overline{\zeta }) &=&\text{\dh }_{(\tau )}X=\xi ^{a}(\tau
)m_{a}(\zeta ,\overline{\zeta })+\text{\dh }_{(\tau )}X_{l\geq 2}(\tau
,\zeta ,\overline{\zeta })  \label{L(u)} \\
u_{B} &=&X=\xi ^{a}(\tau )l_{a}(\zeta ,\overline{\zeta })+X_{l\geq 2}(\tau
,\zeta ,\overline{\zeta }).  \label{u2}
\end{eqnarray}

\begin{remark}
We emphasize that the complex world-line, $\xi ^{a}(\tau ),$ is not in
physical space but is in the parameter space, $H$-space. At this point we
are not suggesting that there is anything profound about this observation.
It is simply there and whatever meaning it might have is obscure. This
observation applies both to the vacuum Einstein equations as well as to the
Einstein-Maxwell equations. Shortly we will see how to make unique choices
of the world-line; in general the Maxwell field will have its own complex
world-line while the Weyl tensor will determine a different world-line. 
\textit{We will consider here only the case where these two world-lines
coincide.}
\end{remark}

\begin{remark}
To invert the latter equation, (\ref{u2}), to obtain $\tau =T(u_{B},\zeta ,%
\overline{\zeta }),$ is, in general, virtually impossible. There however is
a relatively easy method, by iteration, to get approximate inversions to any
accuracy. \{See Appendix A\}\newline
\end{remark}

\begin{remark}
We have tacitly assumed throughout that the original asymptotically flat
space-time metric is real analytic or can be arbitrarily well approximated
by a real analytic metric$.$\newline
\end{remark}

If we expand $\xi ^{a}(\tau )$ in a Taylor series and regroup the terms with
real coefficients and those with imaginary coefficients separately, we can
write 
\begin{equation}
\xi ^{a}(\tau )=\xi _{R}^{a}(\tau )+i\xi _{I}^{a}(\tau ).  \label{R+iI}
\end{equation}

This decomposition becomes important later.\newline

\begin{remark}
The asymptotic twist, $\Sigma (u_{B},\zeta ,\overline{\zeta }),$ \{a measure
of how far the vector field $l^{*a}$ is from being surface forming\} is
defined by 
\begin{equation}
2i\Sigma =\text{\dh }\overline{L}+L\overline{L}^{\cdot }-\overline{\text{\dh 
}}L-\overline{L}L^{\cdot }.
\end{equation}

If $L$ is obtained from a shear that is pure `electric', then the $\xi
^{a}(\tau )$ can be chosen so that the asymptotic twist vanishes.\newline
\end{remark}

It is from the fact that we can chose $L(u_{B},\zeta ,\overline{\zeta })\ $
as a solution of Eq.(\ref{ShearFree}), leading to $\sigma ^{*}=\sigma
^{**}=0,$ that is the justification for the form of Eqs.(\ref{ME1T})-(\ref
{twisting8}).\newline

\subsection{Determining the complex curve.}

The argument and method for the unique determination of the complex curve
(and eventually the definitions of spin and center of mass motion) are not
along conventional lines of thought. To try to clarify our argument, a brief
detour might be worthwhile.

It is well-known that for a static charge distribution, the electric dipole
moment is given by the total charge times the center of charge position
given in the `static' Lorentz frame. In the case of a dynamic charge it is
more difficult since the center of charge depends on the choice of Lorentz
frame. Nevertheless by having, in any one frame, the center of charge at the
coordinate origin, the electric dipole moment will vanish. In the case of a
single charged particle moving on an arbitrary real world line, (the
Lienard-Wiechert Maxwell field) the electric dipole moment vanishes when the
coordinate origin follows the particle's motion or equivalently, the dipole
is given by the particles displacement from the coordinate origin times the
charge. The asymptotically defined dipole moment also vanishes when it is
calculated (or extracted) from the $l=1$ part of the coefficient of the $%
r^{-3}$ term of $\phi _{0}=F_{ab}l^{a}m^{b},$ where $l^{a}$ is a tangent
vector of the light-cones that are attached to the particles world-line.
Analogously, it was shown\cite{ShearFreeMax} that the magnetic dipole moment
of a charged particle can be `viewed' as arising from a charge moving in
complex Minkowski space along a complex world-line. It is given by the
charge times the imaginary displacement. \{We emphasize again that the
complex world-lines are a bookkeeping device and no claim is made that
particles are really moving in complex space. The \textit{real} effect of
the complex world-line picture is to create a twisting real null
congruence.\} If the light-cone from the complex world-line is followed to $%
\frak{I}^{+},$ (equivalent to introducing an appropriate asymptotic twist
for the null vector $l^{a}$ that is used in $\phi _{0}=F_{ab}l^{a}m^{b}$)
then both the asymptotic electric and magnetic dipoles vanish. This argument
is given in considerable detail in the paper\cite{ShearFreeMax,KN}.

The idea was to generalize this construction to GR. In linearized GR and in
the exact case of the Kerr and charged Kerr metrics, the mass-dipole moment
and the spin appear in the real and imaginary parts of the $l=1$ harmonic of
the coefficient of the $r^{-4}$ part of the Weyl component, $\psi
_{1}^{0\,}, $ in a Bondi coordinate/tetrad system. This observation was
extended to all asymptotically flat vacuum solutions by going to an
asymptotically twisting tetrad, with an appropriately chosen $H$-space
complex curve [see below], such that the new Weyl component, $\psi
_{1}^{*\,0\,},$ has a \textit{vanishing} $l=1$ harmonic in the coefficient
of its $r^{-4}$ part. \{Note that for algebraically special type II
space-times the entire quantity $\psi _{1}^{*0\,}=0,$ so by requiring that
the coefficient of the lowest harmonic , $l=1$, should vanish, we are
imitating the type\ II properties as closely as possible.\} We refer to this
curve as the \textit{intrinsic complex center of mass world-line} and,
roughly speaking, it will identify the center of mass and the spin angular
momentum from its real and imaginary parts times the mass. Effectively, by
choosing the world-line so that the $l=1$ harmonic vanishes, we have shifted
- in some sense - the `origin' to the complex world-line.\newline

To apply this idea to our discussion of twisting null congruences for the
Einstein-Maxwell case, we first treat the Maxwell equations. From the pair, (%
\ref{ME1T}) and (\ref{ME2T}), that is 
\begin{eqnarray}
\phi _{0}^{*0\cdot }+\text{\dh }\phi _{1}^{*0}+2L^{\cdot }\phi
_{1}^{*0}+L\phi _{1}^{*0\cdot } &=&0,  \label{M1T*} \\
\phi _{1}^{*0\cdot }+\text{\dh }\phi _{2}^{*0}+(L\phi _{2}^{*0})^{\cdot }
&=&0,  \label{M2T*}
\end{eqnarray}
we require that in (\ref{M1T*}), the $l=1$ part of $\phi _{0}^{*\,0}$ should
vanish. This determines, in principle, the complex world-line, $\xi
^{k}(\tau )$, and leads to the definition of the \textit{complex center of
charge world-line}. \{In practice we must use approximations for this
determination.\} In this work, as was mentioned earlier, we are making,
though it is not necessary, the \textit{special assumption} that the two
complex world-lines, the \textit{complex center of charge world-line} and
the \textit{complex center of mass world-line }[determined later]\textit{\
coincide}.

\begin{remark}
In what follows we use the recently introduced Tensor Spin-s Harmonics%
\textit{\cite{TsH},} $Y_{(l)i...j}^{(s)},$ defined from the tensor products
of the three basic Euclidean vectors $(c_{i},m_{i},\overline{m}_{i})$. See
Appendix C for some notation and gr-qc/0508028 for details. In particular ($%
Y_{1i}^{0}\equiv -c_{i}\equiv -2l_{i}$, $Y_{1i}^{1}\equiv
m_{i},Y_{1i}^{-1}\equiv \overline{m}_{i}).$
\end{remark}

By using the freedom in the reparametrization of $\tau $ given by (\ref
{repar}) and Eq.\thinspace (\ref{X1.2}) in Eqs.\thinspace (\ref{L(u)}) and (%
\ref{u2}) and expanding, we obtain that (see appendix A for details) 
\begin{eqnarray}
&&\tau \,\,\approx \,\,u_{B}-\xi ^{i}(u_{B})l_{i}(\zeta ,\overline{\zeta }
)+X_{l\geq 2}(u_{B},\zeta ,\overline{\zeta }) \\
\xi ^{a}(\tau ) &\equiv &(\xi ^{0}(\tau ),\xi ^{k}(\tau )\,)=(\sqrt{2}\tau
,\xi ^{k}(\tau )\,) \\
&&\xi ^{k}(\tau )\,\,\approx \,\xi ^{k}(u_{B})+\frac{1}{2}\xi ^{i\cdot
}(u_{B})\xi ^{j}(u)Y_{1j}^{0}  \label{approx2}
\end{eqnarray}
and hence from Eq.(\ref{L(u)}) and 
\begin{equation}
Y_{1i}^{1}Y_{1j}^{0}=\frac{i}{\sqrt{2}}\epsilon _{ijk}Y_{1k}^{1}+\frac{1}{2}
Y_{2ij}^{1}.  \label{Y1Y0}
\end{equation}
we find

\begin{equation}
L(u_{B},\zeta ,\overline{\zeta })=[\xi ^{k}(u_{B})+\frac{i}{2\sqrt{2}}
\epsilon _{ijk}\xi ^{i\cdot }(u_{B})\xi ^{j}(u_{B})]Y_{1k}^{1}+Y_{l\geq 2}.
\label{LuF}
\end{equation}

Since $\phi _{1}^{*0}$ and $\phi _{2}^{*0}$ have spin weight $0$ and $-1$
respectively, then they can be written in the following form 
\begin{eqnarray}
\phi _{1}^{*0} &=&C+C^{i}Y_{1i}^{0}+...  \nonumber \\
\phi _{2}^{*0} &=&G^{i}Y_{1i}^{-1}+...  \label{solME}
\end{eqnarray}
For our approximations, (essentially, perturbations off the
Reissner-Nordstrom metric), we choose $C$ as a zero order quantity, and $\xi
^{i}(u_{B})$, $C^{i}(u_{B})$ and $G^{i}(u_{B})$ as first-order. We are
considering only $l=0$ and $1$ terms, i.e., the monopole and dipole.

From Eqs.\thinspace (\ref{LuF}) and (\ref{solME}) we find that 
\begin{eqnarray}
\text{\dh }\phi _{1}^{*0} &=&-2C^{i}Y_{1i}^{1}+...  \nonumber \\
L{\phi }_{1}^{*0\cdot } &=&\{C^{\cdot }[\xi ^{k}+\frac{i}{2\sqrt{2}}\epsilon
_{ijk}\xi ^{i\cdot }\xi ^{j}]+\frac{i}{\sqrt{2}}\epsilon _{ijk}\xi
^{i}C^{j\cdot }\}Y_{1k}^{1}+...  \nonumber \\
L^{\cdot }{\phi }_{1}^{*0} &=&\{C[\xi ^{k\cdot }+\frac{i}{2\sqrt{2}}\epsilon
_{ijk}\xi ^{i\cdot \cdot }\xi ^{j}]+\frac{i}{\sqrt{2}}\epsilon _{ijk}\xi
^{i\cdot }C^{j}\}Y_{1k}^{1}+...,
\end{eqnarray}
where we have used the Eq.(\ref{Y1Y0}).

Therefore, working with the leading terms, we find that the condition $[\phi
_{0}^{*0}]_{l=1}=0$, is equivalent to: 
\begin{eqnarray}
C^{k} &=&\frac{1}{2}C^{\cdot }[\xi ^{k}+\frac{i}{2\sqrt{2}}\epsilon
_{ijk}\xi ^{i\cdot }\xi ^{j}]+\frac{i}{2\sqrt{2}}\epsilon _{ijk}\xi
^{i}C^{j\cdot }  \nonumber \\
&&+C[\xi ^{k\cdot }+\frac{i}{2\sqrt{2}}\epsilon _{ijk}\xi ^{i\cdot \cdot
}\xi ^{j}]+\frac{i}{\sqrt{2}}\epsilon _{ijk}\xi ^{i\cdot }C^{j}.
\label{Cphi*00l=1}
\end{eqnarray}
To see the meaning of this condition we go back to Eq.\thinspace (\ref{M2T*}
). By using Eqs.\thinspace (\ref{LuF}) and (\ref{solME}), a direct
computation shows that: 
\begin{eqnarray}
\text{\dh }\phi _{2}^{*0} &=&G^{i}Y_{1i}^{0}+...,  \nonumber \\
L^{\cdot }\phi _{2}^{*0} &=&\frac{1}{3}\xi ^{j\cdot }G^{j}-\frac{i}{2\sqrt{2}%
}\epsilon _{ijk}\xi ^{i\cdot }G^{j}Y_{1k}^{0}+...,  \nonumber \\
L\phi _{2}^{0\cdot } &=&\frac{1}{3}\xi ^{j}G^{j\cdot }-\frac{i}{2\sqrt{2}}%
\epsilon _{ijk}\xi ^{i}G^{j\cdot }Y_{1k}^{0}]+...,  \label{com2}
\end{eqnarray}
where we have used the product decomposition 
\[
Y_{1k}^{1}Y_{1f}^{-1}=\frac{1}{3}\delta _{kf}-\frac{i}{2\sqrt{2}}\epsilon
_{kfl}Y_{1l}^{0}-\frac{1}{12}Y_{2kf}^{0}. 
\]
Therefore, considering only the leading terms and the $l=0$ and $l=1$
harmonics we find, by using Eqs.\thinspace (\ref{com2}), that Eq.\thinspace (%
\ref{M2T*}) requires 
\begin{eqnarray}
C^{\cdot } &=&-\frac{1}{3}[\xi ^{j}G^{j}]^{\cdot },  \label{Cdot} \\
C^{f\cdot } &=&-G^{f}+\frac{i}{2\sqrt{2}}\epsilon _{ijf}[\xi
^{i}G^{j}]^{\cdot }.  \label{Cfdot}
\end{eqnarray}
If we assume, for the moment, that $\xi ^{i}(\tau )$ and therefore $\xi
^{i}(u_{B})$ is known, then Eqs.\thinspace (\ref{Cphi*00l=1}), (\ref{Cdot})
and (\ref{Cfdot}) are a system of three differential equations for $C$, $%
C^{i}$ and $G^{i}$. Allowing only second order terms of $\xi ^{i}(u)$, the
solution to this system is given by 
\begin{eqnarray}
C &=&q+\frac{q}{3}[\xi ^{j}\xi ^{j\cdot \cdot }]+...,  \nonumber \\
C^{f} &=&q\xi ^{f\cdot }+...,  \nonumber \\
G^{f} &=&-q[\xi ^{f}+\frac{i}{2\sqrt{2}}\epsilon _{ijf}\xi ^{i}\xi ^{j\cdot
}]^{\cdot \cdot }+...,
\end{eqnarray}
where $q$ is a constant. Therefore, the condition that $\phi _{0}^{*0}$
should have no $l=1$ part, implies that the solution of the Maxwell
equations up to $l=1$ is given by 
\begin{eqnarray}
\phi _{0}^{*0} &=&0  \nonumber \\
\phi _{1}^{*0} &=&q+\frac{q}{3}[\xi ^{j}\xi ^{j\cdot \cdot }]+q\xi ^{f\cdot
}Y_{1f}^{0}  \nonumber \\
\phi _{2}^{*0} &=&-q[\xi ^{f\cdot }+\frac{i}{2\sqrt{2}}\epsilon _{ijf}\xi
^{i}\xi ^{j\cdot \cdot }]^{\cdot }Y_{1f}^{-1}.  \label{solME*}
\end{eqnarray}
Observe that $\phi _{2}^{*0}$ can be written as 
\begin{equation}
\phi _{2}^{*0}\equiv \phi _{2}^{0}=-D^{i\cdot \cdot }Y_{1i}^{-1},
\end{equation}
where 
\begin{equation}
D^{f}(u_{B})\equiv q[\xi ^{f}(u_{B})+\frac{i}{2\sqrt{2}}\epsilon _{ijf}\xi
^{i}(u_{B})\xi ^{j\cdot }(u_{B})],  \label{CCM}
\end{equation}
which, up to our order of approximation, is the complex dipole moment. This
means that if we know the complex dipole moment then the complex center of
charge can be obtained. This result had been obtained earlier\cite{KN} for
Maxwell fields in Minkowski space using the following procedure: first the
asymptotic Maxwell equations in a Bondi coordinates/tetrad system were
integrated assuming that the complex dipole moment is given and then by
using the twisting tetrad version of the Maxwell field, $\phi _{0}^{*0},$
the $\xi ^{i}(\tau )$ was determined by requiring that the $l=1$ part of $%
\phi _{0}^{*0}$ should vanish.\newline

Note that in above argument, we have not obtained any dynamics for the
complex curve $\xi ^{i}(\tau );$ all we did was relate it to the electric
and magnetic dipole moments of the Maxwell field. The dynamics is found by
going to the Weyl tensor.

Explicitly we study the asymptotic Bianchi identities under the condition
that the $l=1$ part of $\psi _{1}^{*0}$ should vanish. We retain the
assumption that the $l=1$ of $\phi _{0}^{*0}$ vanish with the same choice of 
$\xi ^{i}(\tau )$, i.e., that the two complex curves coincide.

Starting with the set of equations (\ref{twisting2}) and (\ref{twisting3}),
namely 
\begin{eqnarray}
\psi _{1}^{*0\,\cdot } &=&-\text{\dh }\psi _{2}^{*0}-L\psi _{2}^{*0\,\cdot
}-3L^{\cdot }\psi _{2}^{*0}+2k\phi _{1}^{*0}\overline{\phi }_{2}^{*0}
\label{twisting2*} \\
\psi _{2}^{*0\,\cdot } &=&-\text{\dh }\psi _{3}^{*0}-L\,\psi _{3}^{*0\,\cdot
}-2L^{\cdot }\psi _{3}^{*0}+k\phi _{2}^{*0}\overline{\phi }_{2}^{*0}
\label{twisting3*}
\end{eqnarray}
we require, that in (\ref{twisting2*}), the $l=1$ part of $\psi _{1}^{*0\,}$
should vanish. This leads (via approximations) to differential equations for
the $\xi ^{a}(\tau ).$ Its evolution will be seen to be driven by the
original Bondi shear and the Maxwell field.

Before turning to the dynamics of $\xi ^{a}(\tau )$ we recall the definition
of the Bondi mass-momentum and its relationship to the mass aspect $\Psi $
and its evolution. Eq.(\ref{twisting3*}) can be rewritten as

\begin{eqnarray}
\Psi ^{\cdot } &=&\sigma ^{\cdot }\overline{\sigma }^{\cdot }+k\phi _{2}^{*0}%
\overline{\phi }_{2}^{*0},  \label{1} \\
\sigma &\equiv &\text{\dh }L+LL^{\cdot }  \label{2*} \\
\psi _{2}^{*0} &=&\Psi -2L\text{\dh }\overline{\sigma }^{\cdot }-L^{2}%
\overline{\sigma }^{\cdot \cdot }-\text{\dh }^{2}\overline{\sigma }-\sigma 
\overline{\sigma }^{\cdot }.  \label{3} \\
\phi _{2}^{*0} &=&=-q[\xi ^{f\cdot }+\frac{i}{2\sqrt{2}}\epsilon _{ijf}\xi
^{i}\xi ^{j\cdot \cdot }]^{\cdot }Y_{1f}^{-1}+  \label{4}
\end{eqnarray}

Using the definition\cite{CTG3}[see Appendix B] that the Bondi
four-momentum, $P_{a},$ can be extracted from $\Psi $ by

\begin{equation}
P_{a}(u_{B})\equiv (Mc,-P^{i})=-\frac{c^{3}}{8\pi G}\int \Psi l_{a}dS
\label{P}
\end{equation}
then, by inversion, one can see that

\begin{equation}
\Psi (u_{B},\zeta ,\overline{\zeta })=-\frac{2\sqrt{2}G}{c^{2}}M-\frac{6G}{%
c^{3}}P^{i}Y_{1i}^{0}+\Psi _{l\geq 2}  \label{PSI.II}
\end{equation}
where $\Psi _{l\geq 2}$ contain only harmonics $l\geq 2.$ Since $\Psi $ is
real it follows that both $M$ and $P^{i}$ are real.

The time evolution of $P_{a}(u)$ is found from Eq.(\ref{1}) by integration
over the sphere, i.e., 
\begin{equation}
P_{a}^{\cdot }(u_{B})=-\frac{c^{3}}{8\pi G}\int (\sigma ^{\cdot }\overline{%
\sigma }^{\cdot }+k\phi _{2}^{*0}\overline{\phi }_{2}^{*0})l_{a}dS,
\label{evolution}
\end{equation}
the Bondi mass/momentum loss.

If $\psi _{2}^{*0},$ from Eq.(\ref{3}), is now substituted into (\ref
{twisting2*}) we obtain after using Eq.(\ref{1}) and regrouping, a truly
ugly equation that can be written as

\begin{equation}
\psi _{1}^{*0\,\cdot }+\text{\dh }\Psi +3L^{\cdot }\Psi -\text{\dh }^{3}%
\overline{\sigma }+k[L\phi _{2}^{*0}\overline{\phi }_{2}^{*0}-2\phi _{1}^{*0}%
\overline{\phi }_{2}^{0}]=(NL),  \label{psi0dot}
\end{equation}
with the non-linear terms, (NL), expressed by 
\begin{eqnarray}
(NL) &=&\text{\dh }[\overline{\sigma }^{\cdot }\sigma ]+2\sigma \text{\dh }%
\overline{\sigma }^{\cdot }+3L\text{\dh }^{2}\overline{\sigma }^{\cdot
}+L^{3}\overline{\sigma }^{\cdot \cdot \cdot }+3L\sigma \overline{\sigma }
^{\cdot \cdot }+3L^{2}\text{\dh }\overline{\sigma }^{\cdot \cdot }
\label{NL} \\
&&+3L^{\cdot }[2L\text{\dh }\overline{\sigma }^{\cdot }+L^{2}\overline{
\sigma }^{\cdot \cdot }+\text{\dh }^{2}\overline{\sigma }+\sigma \overline{%
\sigma }^{\cdot }].  \nonumber
\end{eqnarray}

The grouping has the non-linear higher order terms put into (NL) while the
`controllable' (lower order) terms are on the left side. It is from this
equation and (\ref{evolution}) we find the equations of motion. We require,
as said earlier, that $\psi _{1}^{*0\,}$ (and of course $\psi _{1}^{*0\cdot
\,}$) have no $l=1$ harmonics. Since \dh $^{3}\overline{\sigma }$ has spin $%
s=2,$ there is no $l=1$ term and thus we see that Eq.(\ref{psi0dot}) reduces
to the condition that 
\begin{equation}
\lbrack \text{\dh }\Psi +3L^{\cdot }\Psi +k(L\phi _{2}^{*0}\overline{\phi }
_{2}^{*0}-2\phi _{1}^{*0}\overline{\phi }_{2}^{0})]_{l=1}=(NL)_{l=1}.
\label{condition}
\end{equation}
It is this equation that we must analyze, or more accurately approximate, by
looking only at the leading spherical harmonic terms. From Eqs.(\ref{PSI.II}
) and (\ref{LuF}), i.e., from

\begin{eqnarray*}
\Psi (u,\zeta ,\overline{\zeta }) &=&-\frac{2\sqrt{2}G}{c^{2}}M-\frac{6G}{
c^{3}}P^{i}Y_{1i}^{0}+.... \\
L(u,\zeta ,\overline{\zeta }) &=&[\xi ^{k}(u)+\frac{i}{2\sqrt{2}}\epsilon
_{ijk}\xi ^{i\cdot }(u)\xi ^{j}(u)]Y_{1k}^{1}+...
\end{eqnarray*}
it follows that

\begin{eqnarray}
\text{\dh }\Psi &=&12\frac{G}{c^{3}}P^{i}Y_{1i}^{1}  \label{inter} \\
3L^{\cdot }\Psi &=&-\frac{6\sqrt{2}G}{c^{2}}M[\xi ^{k\cdot }+\frac{i}{2\sqrt{%
2}}\epsilon _{ijk}\xi ^{i\cdot \cdot }\xi ^{j}]Y_{1k}^{1}-\frac{18G}{c^{3}}%
\xi ^{i\cdot }P^{j}Y_{1i}^{1}Y_{1j}^{0}  \nonumber \\
&=&-\frac{6\sqrt{2}G}{c^{2}}M[\xi ^{k\cdot }+\frac{i}{2\sqrt{2}}\epsilon
_{ijk}\xi ^{i\cdot \cdot }\xi ^{j}]Y_{1k}^{1}-\frac{18iG}{c^{3}\sqrt{2}}
\epsilon _{ijk}\xi ^{i\cdot }P^{j}Y_{1k}^{1}+Y\{l\geq 2\}  \nonumber \\
\phi _{1}^{*0}\overline{\phi }_{2}^{*0} &=&-q^{2}\{[\overline{\xi }^{k\cdot
\cdot }-\frac{i}{2\sqrt{2}}\epsilon _{ijk}(\overline{\xi }^{i}\overline{\xi }%
^{j\cdot \cdot })^{\cdot }]Y_{1k}^{1}+\overline{\xi }^{i\cdot \cdot }\xi
^{j\cdot }Y_{1i}^{1}Y_{1j}^{0}\}  \nonumber \\
&=&-q^{2}\{\overline{\xi }^{k\cdot \cdot }-\frac{i}{2\sqrt{2}}\epsilon
_{ijk}(\overline{\xi }^{i}\overline{\xi }^{j\cdot \cdot })^{\cdot }+\frac{i}{%
\sqrt{2}}\epsilon _{ijk}\overline{\xi }^{i\cdot \cdot }\xi ^{j\cdot
}\}Y_{1k}^{1}+Y\{l\geq 2\}.
\end{eqnarray}
where we have used Eq.\thinspace (\ref{Y1Y0}). Then, from Eq.(\ref{condition}%
), keeping only the first and second order terms, we have from the $l=1$
harmonic 
\begin{eqnarray}
&&12\frac{G}{c^{3}}P^{k}Y_{1k}^{1}-\frac{6\sqrt{2}G}{c^{2}}M[\xi ^{k\cdot }+%
\frac{i}{2\sqrt{2}}\epsilon _{ijk}\xi ^{i\cdot \cdot }\xi ^{j}]Y_{1k}^{1}-%
\frac{18iG}{c^{3}\sqrt{2}}\epsilon _{ijk}\xi ^{i\cdot }P^{j}Y_{1k}^{1} 
\nonumber \\
&&+\frac{4Gq^{2}}{c^{4}}\{\overline{\xi }^{k\cdot \cdot }-\frac{i}{2\sqrt{2}}%
\epsilon _{ijk}[(\overline{\xi }^{i}\overline{\xi }^{j\cdot \cdot })^{\cdot
}-2\overline{\xi }^{i\cdot \cdot }\xi ^{j\cdot }]\}Y_{1k}^{1}=0,
\end{eqnarray}
our basic equation which determines the Bondi momentum in terms of the
complex world-line.

Rescaling $u_{B}$ to the conventional Bondi coordinate$^{1}$ $u_{c}$ , by $%
u_{B}=\frac{1}{\sqrt{2}}cu_{c}$, we have our basic equation, the Bondi
momentum in terms of the complex world-line; 
\begin{eqnarray}
&&P^{k}=M[\xi ^{k\cdot }+\frac{i}{2c}\epsilon _{ijk}\xi ^{i\cdot \cdot }\xi
^{j}]+\frac{3i}{2c}\epsilon _{ijk}\xi ^{i\cdot }P^{j}  \nonumber \\
&&-\frac{2q^{2}}{3c^{3}}\{\overline{\xi }^{k\cdot \cdot }-\frac{i}{2c}%
\epsilon _{ijk}[(\overline{\xi }^{i}\overline{\xi }^{j\cdot \cdot })^{\cdot
}-2\overline{\xi }^{i\cdot \cdot }\xi ^{j\cdot }]\}.  \label{Pk***}
\end{eqnarray}

\begin{description}
\item  footnote$^{1}:$ Many years ago, for reasons of symmetry and
convenience the conventional radial coordinate $r_{c}$ was rescaled by a
factor of $\sqrt{2},$ i.e., $r=\sqrt{2}r_{c}$ , so that to keep a coordinate
condition the retarded time coordinate $u_{c}$ was rescaled so that our $%
u_{B}=\frac{1}{\sqrt{2}}u_{c}.$ Here we will restore the conventional $u_{c}$
but also include the ordinary time units \{i.e., with c$\neq 1$\} we take $%
u_{B}=\frac{1}{\sqrt{2}}cu_{c},$ so that derivatives $\partial _{u_{B}}=%
\frac{\sqrt{2}}{c}\partial _{u_{c}}.$
\end{description}

By using the decomposition 
\[
\xi ^{k}=\xi _{R}^{k}+i\xi _{I}^{k}, 
\]
and then iterating, i.e., by substituting into the right-side of Eq.(\ref
{Pk***} ) the real linearized expression for $P^{k},$%
\[
P^{k}=M\xi _{R}^{k\cdot }-\frac{2q^{2}}{3c^{3}}\xi _{R}^{k\cdot \cdot }, 
\]
into (\ref{Pk***}), we find that the Real Part of Eq. (\ref{Pk***}) is given
by:

\begin{eqnarray}
P^{k} &=&M[\xi _{R}^{k\cdot }-\frac{3}{2c}\epsilon _{ijk}\xi _{I}^{i\cdot
}\xi _{R}^{j\cdot }-\frac{1}{2c}\epsilon _{ijk}(\xi _{I}^{i\cdot \cdot }\xi
_{R}^{j}+\xi _{R}^{i\cdot \cdot }\xi _{I}^{j})]-\frac{2q^{2}}{3c^{3}}\xi
_{R}^{k\cdot \cdot }  \nonumber \\
&&+\frac{q^{2}}{3c^{4}}\epsilon _{ijk}\{6\xi _{I}^{i\cdot }\xi _{R}^{j\cdot
\cdot }-\xi _{I}^{j\cdot \cdot }\xi _{R}^{i\cdot }+\xi _{I}^{i}\xi
_{R}^{j\cdot \cdot \cdot }+\xi _{R}^{i}\xi _{I}^{j\cdot \cdot \cdot }\},
\label{Pk&}
\end{eqnarray}
while the Imaginary Part is 
\begin{eqnarray}
&&0=M\xi _{I}^{k\cdot }+\frac{M}{2c}\epsilon _{ijk}[\xi _{R}^{i\cdot \cdot
}\xi _{R}^{j}-\xi _{I}^{i\cdot \cdot }\xi _{I}^{j}]+\frac{2q^{2}}{3c^{3}}\xi
_{I}^{k\cdot \cdot }  \label{IMAGPk} \\
&&+\frac{q^{2}}{3c^{4}}\epsilon _{ijk}[\xi _{R}^{i}\xi _{R}^{j\cdot \cdot
\cdot }-\xi _{I}^{i}\xi _{I}^{j\cdot \cdot \cdot }-\xi _{I}^{i\cdot \cdot
}\xi _{I}^{j\cdot }].  \nonumber
\end{eqnarray}

We can now obtain the equations of motion for $\xi _{R}^{k}$ by taking the $%
u_{B}$ derivative of Eq.(\ref{Pk&}), i.e.,

\begin{eqnarray}
P^{k\cdot } &=&M^{\cdot }[\xi _{R}^{k\cdot }-\frac{3}{2c}\epsilon _{ijk}\xi
_{I}^{i\cdot }\xi _{R}^{j\cdot }-\frac{1}{2c}\epsilon _{ijk}(\xi
_{I}^{i\cdot \cdot }\xi _{R}^{j}+\xi _{R}^{i\cdot \cdot }\xi _{I}^{j})]-%
\frac{2q^{2}}{3c^{3}}\xi _{R}^{k\cdot \cdot \cdot }  \nonumber \\
&&+M[\xi _{R}^{k\cdot \cdot }-\frac{3}{2c}\epsilon _{ijk}(\xi _{I}^{i\cdot
}\xi _{R}^{j\cdot })^{\cdot }-\frac{1}{2c}\epsilon _{ijk}(\xi _{I}^{i\cdot
\cdot }\xi _{R}^{j}+\xi _{R}^{i\cdot \cdot }\xi _{I}^{j})^{\cdot }] 
\nonumber \\
&&+\frac{q^{2}}{3c^{4}}\epsilon _{ijk}\{6\xi _{I}^{i\cdot }\xi _{R}^{j\cdot
\cdot }-\xi _{I}^{j\cdot \cdot }\xi _{R}^{i\cdot }+\xi _{I}^{i}\xi
_{R}^{j\cdot \cdot \cdot }+\xi _{R}^{i}\xi _{I}^{j\cdot \cdot \cdot
}\}^{\cdot }.  \label{PkdotI}
\end{eqnarray}
and equate the right side to the Bondi mass, momentum loss, i.e., to Eq.(\ref
{evolution}) 
\begin{eqnarray}
P_{a}^{\cdot }(u_{B}) &=&-\frac{c^{3}}{8\pi G}\int (\sigma ^{\cdot }%
\overline{\sigma }^{\cdot }+k\phi _{2}^{*0}\overline{\phi }_{2}^{*0})l_{a}dS.
\label{evolution2} \\
k &=&\frac{2G}{c^{4}}
\end{eqnarray}

\{Note that care must be exercised when using this equation since all the $%
\partial _{u_{B}}$ derivatives must be replaced by the rescaled derivatives $%
\frac{\sqrt{2}}{c}\partial _{u_{B}}.\}$

In order to evaluate the integrand and the integral, we assume only the
lowest gravitational harmonics, i.e., the quadrupole radiation and hence
take 
\begin{equation}
\sigma =\sigma ^{ij}(u_{B})Y_{2ij}^{2}+Y_{l\geq 3},  \label{sigma*}
\end{equation}
where $Y_{2ij}^{2}\equiv m_{i}m_{j}$. Then 
\begin{equation}
\sigma ^{\cdot }\overline{\sigma }^{\cdot }=\sigma ^{ij\cdot }\overline{
\sigma }^{kl\cdot }Y_{2ij}^{2}Y_{2kl}^{-2}+....,
\end{equation}
with $Y_{2kl}^{-2}\equiv \overline{Y_{2ij}^{2}}.$ After a direct but lengthy
computation\cite{TsH} to evaluate the products of the harmonics, we find that

\begin{equation}
\sigma ^{\cdot }\overline{\sigma }^{\cdot }=\frac{1}{5}\overline{\sigma }%
^{ij \cdot }\sigma ^{ij\cdot }+i\frac{\sqrt{2}}{5}\overline{\sigma }%
^{il\cdot }\sigma ^{jl\cdot }\epsilon _{ijk}Y_{1k}^{0}+Y_{l\geq 2}.
\label{sigmadot^2}
\end{equation}

Likewise we have that 
\begin{equation}
\phi _{2}^{*0}\overline{\phi }_{2}^{*0}=\frac{q^{2}}{3}\overline{\xi }^{i
\cdot \cdot }\xi ^{i\cdot \cdot }-\frac{iq^{2}}{2\sqrt{2}}\epsilon _{ijk} 
\overline{\xi }^{i\cdot \cdot }\xi ^{j\cdot \cdot }Y_{1k}^{0}+Y_{l\geq 2}.
\label{phi2^2}
\end{equation}

Putting Eqs.(\ref{phi2^2}) and (\ref{sigmadot^2}) into (\ref{evolution2})
the evolution for the mass and the three momentum is then given by

\begin{eqnarray}
M^{\cdot } &=&-\frac{2q^{2}}{3c^{5}}\overline{\xi }^{i\cdot \cdot }\xi
^{i\cdot \cdot }-\frac{c}{10G}\overline{\sigma }^{ij\cdot }\sigma ^{ij\cdot
},  \nonumber \\
&=&-\frac{2q^{2}}{3c^{5}}[{\xi }_{R}^{i\cdot \cdot }{\xi }_{R}^{i\cdot \cdot
}+{\xi }_{I}^{i\cdot \cdot }{\xi }_{I}^{i\cdot \cdot }]-\frac{c}{10G}%
\overline{\sigma }^{ij\cdot }\sigma ^{ij\cdot },  \label{Mdot} \\
P^{k\cdot } &=&\frac{iq^{2}}{3c^{4}}\epsilon _{ijk}\overline{\xi }^{i\cdot
\cdot }\xi ^{j\cdot \cdot }-\frac{ic^{2}}{15G}\epsilon _{ijk}\overline{%
\sigma }^{il\cdot }\sigma ^{jl\cdot }  \nonumber \\
&=&-\frac{q^{2}}{3c^{4}}\epsilon _{ijk}[{\xi }_{R}^{i\cdot \cdot }{\xi }
_{I}^{j\cdot \cdot }-{\xi }_{R}^{j\cdot \cdot }{\xi }_{I}^{i\cdot \cdot }]-%
\frac{ic^{2}}{15G}\epsilon _{ijk}\overline{\sigma }^{il\cdot }\sigma
^{jl\cdot }  \label{PKdot}
\end{eqnarray}

Using $\xi ^{k}=\xi _{R}^{k}+i\xi _{I}^{k}$ in Eq.\thinspace (\ref{PKdot})
and substituting the resulting equation into Eq.\thinspace (\ref{PkdotI}),
we obtain

\begin{eqnarray}
M\xi _{R}^{k\cdot \cdot } &=&\frac{2q^{2}}{3c^{3}}\xi _{R}^{k\cdot \cdot
\cdot }+M[\frac{3}{2c}\epsilon _{ijk}(\xi _{I}^{i\cdot }\xi _{R}^{j\cdot
})^{ \cdot }+\frac{1}{2c}\epsilon _{ijk}(\xi _{I}^{i\cdot \cdot }\xi
_{R}^{j}+\xi _{R}^{i\cdot \cdot }\xi _{I}^{j})^{\cdot }]  \nonumber \\
&&-\frac{q^{2}}{3c^{4}}\epsilon _{ijk}[{\xi }_{R}^{i\cdot \cdot }{\xi }
_{I}^{j\cdot \cdot }-{\xi }_{R}^{j\cdot \cdot }{\xi }_{I}^{i\cdot \cdot }]- 
\frac{ic^{2}}{15G}\epsilon _{ijk}\overline{\sigma }^{il\cdot }\sigma ^{jl
\cdot }  \nonumber \\
&&-M^{\cdot }[\xi _{R}^{k\cdot }-\frac{3}{2c}\epsilon _{ijk}\xi _{I}^{i\cdot
}\xi _{R}^{j\cdot }-\frac{1}{2c}\epsilon _{ijk}(\xi _{I}^{i\cdot \cdot }\xi
_{R}^{j}+\xi _{R}^{i\cdot \cdot }\xi _{I}^{j})]  \nonumber \\
&&-\frac{q^{2}}{3c^{4}}\epsilon _{ijk}[6\xi _{I}^{i\cdot }\xi _{R}^{j\cdot
\cdot }-\xi _{I}^{j\cdot \cdot }\xi _{R}^{i\cdot }+\xi _{I}^{i}\xi _{R}^{j
\cdot \cdot \cdot }+\xi _{R}^{i}\xi _{I}^{j\cdot \cdot \cdot }]^{\cdot }.
\label{Emotion}
\end{eqnarray}
which after a lengthy calculation becomes 
\begin{eqnarray}
M\xi _{R}^{k\cdot \cdot } &=&\frac{2q^{2}}{3c^{3}}\xi _{R}^{k\cdot \cdot
\cdot }+\frac{M}{2c}\epsilon _{ijk}[3\xi _{I}^{i\cdot }\xi _{R}^{j\cdot
}+\xi _{I}^{i\cdot \cdot }\xi _{R}^{j}+\xi _{R}^{i\cdot \cdot }\xi
_{I}^{j}]^{\cdot }  \nonumber \\
&&-\frac{q^{2}}{3c^{4}}\epsilon _{ijk}[5\xi _{I}^{i\cdot \cdot }\xi _{R}^{j
\cdot \cdot }+7\xi _{I}^{i\cdot }\xi _{R}^{j\cdot \cdot \cdot }+\xi
_{I}^{i}\xi _{R}^{j\cdot \cdot \cdot \cdot }+\xi _{R}^{i}\xi _{I}^{j\cdot
\cdot \cdot \cdot }]  \nonumber \\
&&-M^{\cdot }\xi _{R}^{k\cdot }+\frac{M^{\cdot }}{2c}\epsilon _{ijk}[3\xi
_{I}^{i\cdot }\xi _{R}^{j\cdot }+\xi _{I}^{i\cdot \cdot }\xi _{R}^{j}+\xi
_{R}^{i\cdot \cdot }\xi _{I}^{j}]  \nonumber \\
&&-\frac{ic^{2}}{15G}\epsilon _{ijk}\overline{\sigma }^{il\cdot }\sigma ^{jl
\cdot }.  \label{Emotion***}
\end{eqnarray}

Finally, using for $M^{\cdot }$ the Bondi mass loss equation 
\[
M^{\cdot }=-\frac{2q^{2}}{3c^{5}}[{\xi }_{R}^{i\cdot \cdot }{\xi }_{R}^{i
\cdot \cdot }+{\xi }_{I}^{i\cdot \cdot }{\xi }_{I}^{i\cdot \cdot }]-\frac{c}{
10G}\overline{\sigma }^{ij\cdot }\sigma ^{ij\cdot }, 
\]
in Eq. (\ref{Emotion***}) and replacing $M$ by $M_{0}=constant$ , the zero
order of the Bondi mass we obtain, the equations of motion:

\begin{eqnarray}
M_{0}\xi _{R}^{k\cdot \cdot } &=&\frac{2q^{2}}{3c^{3}}\xi _{R}^{k\cdot \cdot
\cdot }+\frac{M_{0}}{2c}\epsilon _{ijk}[3\xi _{I}^{i\cdot }\xi _{R}^{j\cdot
}+\xi _{I}^{i\cdot \cdot }\xi _{R}^{j}+\xi _{R}^{i\cdot \cdot }\xi
_{I}^{j}]^{\cdot }  \nonumber \\
&&-\frac{q^{2}}{3c^{4}}\epsilon _{ijk}[5\xi _{I}^{i\cdot \cdot }\xi
_{R}^{j\cdot \cdot }+7\xi _{I}^{i\cdot }\xi _{R}^{j\cdot \cdot \cdot }+\xi
_{I}^{i}\xi _{R}^{j\cdot \cdot \cdot \cdot }+\xi _{R}^{i}\xi _{I}^{j\cdot
\cdot \cdot \cdot }]  \nonumber \\
&&+\frac{2q^{2}}{3c^{5}}[{\xi }_{R}^{i\cdot \cdot }{\xi }_{R}^{i\cdot \cdot
}+{\xi }_{I}^{i\cdot \cdot }{\xi }_{I}^{i\cdot \cdot }]\xi _{R}^{k\cdot }-%
\frac{ic^{2}}{15G}\epsilon _{ijk}\overline{\sigma }^{il\cdot }\sigma
^{jl\cdot }.  \label{Emotion****}
\end{eqnarray}

In the last computation we dropped many higher order terms.

For the Imaginary part, replacing $M$ by $M_{0}$ we obtain

\begin{eqnarray}
M_{0}\xi _{I}^{k\cdot } &=&-\frac{M_{0}}{2c}\epsilon _{ijk}[\xi _{R}^{i\cdot
\cdot }\xi _{R}^{j}-\xi _{I}^{i\cdot \cdot }\xi _{I}^{j}]-\frac{2q^{2}}{%
3c^{3}}\xi _{I}^{k\cdot \cdot }  \label{spinevolution} \\
&&-\frac{q^{2}}{3c^{4}}\epsilon _{ijk}[\xi _{R}^{i}\xi _{R}^{j\cdot \cdot
\cdot }-\xi _{I}^{i}\xi _{I}^{j\cdot \cdot \cdot }-\xi _{I}^{i\cdot \cdot
}\xi _{I}^{j\cdot }].  \nonumber
\end{eqnarray}

Before we discuss the meaning of Eqs.(\ref{Emotion****}) and (\ref
{spinevolution}) in the following section, we turn to the special case of $%
\xi _{I}^{j}=0$ and $\overline{\sigma }^{il}=\sigma ^{il}$. The equations of
motion then reduce to

\begin{equation}
M_{0}\xi _{R}^{k\cdot \cdot }=\frac{2q^{2}}{3c^{3}}\xi _{R}^{k\cdot \cdot
\cdot }+[\frac{2q^{2}}{3c^{5}}{\xi }_{R}^{i\cdot \cdot }{\xi }_{R}^{i\cdot
\cdot }+\frac{c}{10G}\sigma ^{ij\cdot }\sigma ^{ij\cdot }]\xi _{R}^{k\cdot },
\label{R.eqs}
\end{equation}
and the Imaginary part

\begin{equation}
\frac{M_{0}}{2c}\epsilon _{ijk}[\xi _{R}^{i\cdot \cdot }\xi _{R}^{j}]+\frac{%
q^{2}}{3c^{4}}\epsilon _{ijk}[\xi _{R}^{i}\xi _{R}^{j\cdot \cdot \cdot }]=0
\label{i.eqs}
\end{equation}
which, using Eq.(\ref{R.eqs}) up to third order, is the identity 
\[
\epsilon _{ijk}\xi _{R}^{j}[M_{0}\xi _{R}^{i\cdot \cdot }-\frac{2q^{2}}{%
3c^{3}}\xi _{R}^{i\cdot \cdot \cdot }]\equiv 0. 
\]

As an aside we note that the complex curve in H-space that we wish to
determine is given by the expression, $z^{a}=\xi ^{b}(\tau ),$ with the
complex parameter $\tau ,$ while in the above expressions for the
determination of the curve we have treated $\xi ^{b}$ to be a complex
function of the retarded time $u_{B},$ which can take on complex values but
we have treated as real. In \textit{terms} of their functional form, $\xi
^{b}(\tau )$ is identical to $\xi ^{b}(u_{B})$, but in terms of the usage in
different equations, the transition via $\tau =T(u_{B},\zeta ,\overline{%
\zeta })$ and its linearized approximation (done in Appendix A) is different
and very important. We have, directly, by this approximation been throwing
out higher order non-linear terms.

\section{Interpretations, Discussion and Conclusions}

The basic idea we are espousing is that there is, in $H$-Space, a unique
complex curve that can be called the intrinsic complex center of mass
world-line and complex center of charge line. It determines and also is
determined from \textit{a geometric structure on the physical space-time,
the null direction field, }which\textit{\ is }given either by the angle field%
\textit{\ }$L(u_{B},\zeta ,\overline{\zeta })$ or by the tangent field of
the asymptotically shear-free null congruence $l^{*a}.$ They are geometric
quantities and exist in the physical space-time independent of the choice of
coordinates or tetrad on $\frak{I}^{+}.$ However the complex curve has no
immediate physical meaning. There are however a series of observations that
do suggest a physical interpretation. Furthermore there are many clues in
earlier literature\cite{PM,N,LN,TedDublin,gyro,QKerr} to its significance.

The basic idea is not to take the parameter space, H-space, as real or
concrete in any sense - but instead treat it as an observation or
holographic space. In the physical space-time we are dealing with a
complicated physical system, a gravitating - charged mass distribution, that
creates curvature and an electromagnetic field that we can `see' or observe
only from its asymptotic behavior; we can not see individual masses,
charges, spins or particle trajectories but instead only the gross or
large-scale behavior. The idea is that the H-space is like a screen that
captures certain images of the physical space. This idea arises from several
directions and, in turn, suggests other physical interpretations.

1. We know how to define energy and momentum from the classical treatment of
the Bondi mass and momentum. Now from the H-space point of view and at a low
order of approximation we have recovered the trivial relationship between
three-momentum, mass and velocity, Eq.(\ref{Pk&}), i.e., $P=M\xi _{R}^{\cdot
}.$

2. In classical Electromagnetic\ theory, using the electromagnetic
self-force and the Lorentz force law, one finds the radiation reaction term $%
\frac{2q^{2}}{3c^{3}}\xi _{R}^{k\cdot \cdot \cdot }.$ Here, just by looking
at linear order, at the asymptotic Einstein-Maxwell fields with no attempt
at model building, we obtain exactly the same term in the equations of
motion without any type of model building. Furthermore, by looking at the
equations of motion for the special case of $\xi _{I}^{k}=0,$ i.e., Eq.(\ref
{R.eqs}), we see that in addition to the radiation reaction term there are
further non-linear terms that arises from the Bondi mass loss formula, Eq.(%
\ref{Mdot}). In simple numerical examples, neglecting the ($\sigma ^{\cdot
})^{2}$ term, it appears as if the classical run-away solutions of just the
radiation reaction term are damped out.

3. In our equation for mass loss, Eq.(\ref{Mdot}), there is electric and
magnetic dipole radiation with the dipole with the second time derivative of
the complex $q\xi ^{k}(u_{B})=q(\xi _{R}^{k}+i\xi _{I}^{k}),$ i.e., we see
that there is energy loss by both electric and magnetic dipole radiation
with the identification of $q\xi _{R}^{k}\ $and $q\xi _{I}^{k}$ as the
electric and magnetic dipole moments. The numerical factors even agree with
dipole radiation derived from pure Maxwell theory in Minkowski space.

It appears as if the $\xi _{R}^{k}(u_{B})$ can be used as a `position
vector' and it is suggestive that $M\xi _{R}^{k}(u_{B})$ and $q\xi
_{R}^{k}(u_{B})$ can be interpreted as the mass and charge dipole moments
and thus $\xi _{R}^{k}(u_{B})$ interpreted as the center of mass and charge.

From knowledge of the Schwarzschild, Kerr, Reissner-Nordstrom and charged
Kerr metrics, one can try to give meaning to the complex curve and in
particular to the \textit{imaginary} part of the curve. It has been well
know since the 1960's that the Kerr space-time has associated with it a
complex space and a unique complex world-line such that its imaginary part, 
\TEXTsymbol{\vert}$\xi _{I}^{k}|=a,$ determines the spin angular-momentum $%
S, $ via $S=Mca.$ Furthermore for the charged Kerr metric there is the 
\textit{same} complex world-line with imaginary part, \TEXTsymbol{\vert}$\xi
_{I}^{k}|=a$ from which both the spin $S$ and magnetic dipole moment, $\mu ,$
are obtained from $S=Mca$ and $\mu =qa.$ In none of these cases is there a
non-vanishing real part of the world-line and hence the real center of mass
and the electric dipole vanished and, hence, are at the origin.

4. These observations suggest that in the present work the imaginary part of
the complex curve, i.e., $\xi _{I}^{k},$ should be used to define both the
spin and the magnetic dipole moment via 
\begin{eqnarray*}
\overrightarrow{S} &=&Mc\overrightarrow{\xi }_{I} \\
\overrightarrow{\mu } &=&q\overrightarrow{\xi }_{I}.
\end{eqnarray*}

5. From a study\cite{gyro,KN,ShearFreeMax} of the Maxwell equations in
Minkowski space, where one needs far less approximations and sometimes no
approximations at all, again we have that $\overrightarrow{\mu }=q%
\overrightarrow{\xi }_{I}.$

6. Though at the present time we do not know the meaning of many of the
terms in both the expression for the 3-momentum, (\ref{Pk&}) and the
equations of motion, (\ref{Emotion****}), it appears possibly that they come
from the fact that we are dealing with the matter from entire interior of
the space-time and information from earlier times is being fed in via the
higher time derivatives. Nevertheless some of the terms are familiar: In (%
\ref{Pk&}) there are three terms

\[
P^{k}=M\xi _{R}^{k\cdot }-\frac{2q^{2}}{3c^{3}}\xi _{R}^{k\cdot \cdot }-%
\frac{3}{2c}M\epsilon _{ijk}\xi _{I}^{i\cdot }\xi _{R}^{j\cdot }+..... 
\]
that are known from other considerations: there is the standard $Mv$ term,
the second term comes from radiation reaction [its time derivative is the
well-known radiation reaction force.] The last term is the spin-velocity
coupling term found in the Mathisson-Papapetrou equations\cite{M,P} if again
we identify the spin with $Mc\xi _{I}^{i\cdot }.$

It thus appears natural, in our case where our two world-lines coincide, to
identify $\xi _{I}^{k}$ as the imaginary center of mass and charge and thus
define the full complex $\xi ^{k}(u_{B})$ as the complex center of mass and
center of charge. This point of view allows a unification of many examples
of electric and magnetic phenomena.

Returning to the charged Kerr case, it was noted that in the ratio $S$/$\mu
=Mc/q,$ the imaginary displacement,$`a$', drops out, and one discovers that
for the charged Kerr solution the gyromagnetic ratio has the Dirac value$,$
i.e., $g=2.$ This result came from the fact that the two complex world-lines
coincided. From this observation we see that, for all asymptotically flat
Einstein-Maxwell fields where the two complex H-space world-lines coincide,
again the gyromagnetic ratio will have the Dirac value, $g=2.$

Though there is still much to understand and do, we believe that there is
something essentially correct in this screen or holographic point of view
towards these H-space curves. A variety of questions, however, do arise.

a. Is this discovery of the complex curve on H-space and the direction
fields just a booking keeping device or is there something deeper. We do not
know but we can see a beautiful symmetry between orbital and spin angular
momentum and between electric and magnetic dipole moments. They are
naturally unified into the complex vector $\xi ^{a}$. This structure takes
on even more beauty in the case of the twisting type II metrics (in some
sense fundamental solutions) where the asymptotically shear-free congruence
is totally shear-free.

b. We do not yet know the connection between our definition of spin and the
more usual one obtained from the symmetry properties of the BMS group.

c. We need to analyze the meaning of many of the terms in the equations of
motion. It appears clear that they will involve couplings between all the
different types of moments; mass-dipole/orbital angular momentum, spin,
electric and magnetic dipoles and higher moments. Will they have any obvious
meaning?

d. There is the issue to be explored of whether it is possible in principle
[almost certainly not in practice] to \textit{observe} the complex
world-line. The answer very likely is yes; it probably can be done by a
spherical harmonic analysis of the angle field, $L(u_{B},\zeta ,\overline{%
\zeta })$, then looking at the $l=1$ component.

e. The potential damping of the classical run-away solution from the
radiation reaction term must be further studied.

f. What relationship does our unique choice of the angle field $%
L(u_{B},\zeta ,\overline{\zeta })$ have to the same geometric object in the
algebraically special metrics where it is used to define a CR structure on $%
\frak{I}^{+}?$ It would appear that a mathematical statement of our results
could be the following: (i). For all asymptotically flat vacuum space-times
there is a unique CR structure on $\frak{I}^{+}$,( ii). For all
asymptotically flat Einstein-Maxwell space-times there are two unique CR
structures on $\frak{I}^{+},$ (iii). In the degenerate case, when the two
structures coincide, one find that the gyromagnetic ratio is that of Dirac,
namely $g=2.$

\section{Acknowledgments}

This material is based upon work (partially) supported by the National
Science Foundation under Grant No. PHY-0244513. Any opinions, findings, and
conclusions or recommendations expressed in this material are those of the
authors and do not necessarily reflect the views of the National Science.
E.T.N. thanks the NSF for this support. G.S.O. acknowledges the financial
support from CONACyT through Grand No.44515-F, VIEP-BUAP through Grant
No.17/EXC/05 and Sistema Nacional de Investigadores (SNI-M\'{e}xico).

\section{Appendix A}

In order to obtain $\tau =T(u_{B},\zeta ,\overline{\zeta }),$ the inversion
of 
\[
u_{B}=X(\tau ,\zeta ,\overline{\zeta })=\xi ^{a}(\tau )l_{a}(\zeta ,%
\overline{\zeta })+X_{l\geq 2}(\tau ,\zeta ,\overline{\zeta }), 
\]
we first note that $\tau $ can be replaced by any function of $\tau ,$ i.e.,
the Eq.(\ref{L1}) is invariant under $\widehat{\tau }=F(\tau ).$ See Eq.(\ref
{repar}). Thus by taking 
\[
\xi ^{0}(\tau )=\sqrt{2}\tau 
\]
we have 
\[
u_{B}=\tau +\xi ^{i}(\tau )l_{i}(\zeta ,\overline{\zeta })+X_{l\geq 2}(\tau
,\zeta ,\overline{\zeta }) 
\]
or 
\[
\tau =u_{B}-\xi ^{i}(\tau )l_{i}(\zeta ,\overline{\zeta })+X_{l\geq 2}(\tau
,\zeta ,\overline{\zeta }). 
\]
This can be solved by iteration to any order: 
\begin{eqnarray*}
\tau _{0} &=&u_{B} \\
\tau _{1} &=&u_{B}-\xi ^{i}(u_{B})l_{i}(\zeta ,\overline{\zeta })+X_{l\geq
2}(u,\zeta ,\overline{\zeta }) \\
\tau _{n} &=&u_{B}-\xi ^{i}(\tau _{n-1})l_{i}(\zeta ,\overline{\zeta }%
)+X_{l\geq 2}(\tau _{n-1},\zeta ,\overline{\zeta }).
\end{eqnarray*}

In the text we have truncated this procedure and used the approximate
inversion as 
\[
\tau =u_{B}-\xi ^{i}(u_{B})l_{i}(\zeta ,\overline{\zeta }). 
\]

\section{Appendix B}

Normalization and conventions: Our goal here is to express the function $%
\Psi $ in terms of the Bondi mass/momenta and also show how to extract them
from the $\Psi .$

We start, in a given frame in Minkowski space, with $C_{a}$ a unit radial
space-like vector and $T_{a}$ a unit time-like vector given by

\begin{eqnarray}
C^{a} &=&(0,\cos \phi \sin \theta ,\sin \phi \sin \theta ,\cos \theta
)=-C_{a}, \\
T_{a} &=&(1,0,0,0).
\end{eqnarray}
with the null vectors 
\begin{eqnarray}
l_{a} &=&\frac{1}{\sqrt{2}}\left( 1,-\cos \phi \sin \theta ,-\sin \phi \sin
\theta ,-\cos \theta \right) =\frac{1}{\sqrt{2}}(T_{a}+C_{a}) \\
n_{a} &=&\frac{1}{\sqrt{2}}\left( 1,\cos \phi \sin \theta ,\sin \phi \sin
\theta ,\cos \theta \right) =\frac{1}{\sqrt{2}}(T_{a}-C_{a})
\end{eqnarray}
and 
\[
c_{a}=l_{a}-n_{a}=-\sqrt{2}(0,\cos \phi \sin \theta ,\sin \phi \sin \theta
,\cos \theta )=\sqrt{2}C_{a} 
\]

We let $a=(0,1,2,3)=(0,i)$ and $l_{a}=(l_{0},l_{i})$ and $c_{i}=\sqrt{2}
C_{i} $, etc.

We then define the $l=0$ and $l=1$ parts of the mass aspect $\Psi $ by

\begin{eqnarray}
\Psi &\equiv &\psi _{2}^{*0}+2L\text{\dh }\overline{\sigma }^{\cdot }+L^{2}%
\overline{\sigma }^{\cdot \cdot }+\text{\dh }^{2}\overline{\sigma }+\sigma 
\overline{\sigma }^{\cdot }  \label{PSI1} \\
\Psi &=&\chi -\chi ^{i}c_{i}+....  \label{PSI2}
\end{eqnarray}
Therefore 
\begin{eqnarray}
\Psi l_{a} &=&\chi l_{a}-\chi ^{i}c_{i}l_{a}+...  \label{PSIL} \\
&=&\frac{1}{\sqrt{2}}[\chi T_{a}-\sqrt{2}\chi ^{i}C_{i}C_{a}+\chi C_{a}-%
\sqrt{2}\chi ^{i}C_{i}T_{a}].....  \nonumber
\end{eqnarray}
From the integral identities, with $dS$ the area element on the unit metric
sphere, 
\begin{eqnarray*}
\int dS &=&4\pi \\
\int C_{i}dS &=&0 \\
\int C_{i}C_{j}dS &=&\frac{4\pi }{3}\delta _{ij}
\end{eqnarray*}
we have that

\begin{eqnarray}
\int \Psi l_{a}dS &=&\frac{1}{\sqrt{2}}\int dS(\chi T_{a}-\sqrt{2}\chi
^{i}C_{i}C_{a})  \nonumber \\
&=&\frac{4\pi }{\sqrt{2}}[\chi T_{a}-\chi ^{i}\frac{\sqrt{2}}{3}]=\frac{4\pi 
}{\sqrt{2}}\left( \chi ,\chi _{i}\frac{\sqrt{2}}{3}\right) .  \label{intPsi}
\end{eqnarray}
with $\chi _{i}=-\chi ^{i}.$\newline

On the other hand, from reference (\cite{TN})\textit{\ }and a change of%
\textit{\ }notation [see footnote 1],\textit{\ \ }we have that the
4-momentum for an asymptotically flat space-time is defined by

\begin{eqnarray}
P_{a} &\equiv &(Mc,-P^{i})=-\frac{c^{3}}{8\pi G}\int \Psi l_{a}dS,
\label{P_a} \\
\Psi &=&-2\sqrt{2}\frac{G}{c^{2}}M+\frac{6G}{c^{3}}P^{i}c_{i}+\Psi _{l\geq
2}.....  \label{PSI*}
\end{eqnarray}

then from Eqs.(\ref{P_a}),(\ref{PSI*}) and (\ref{PSIL}),

\begin{eqnarray}
P_{a} &=&-\frac{c^{3}}{2\sqrt{2}G}\left( \chi ,\chi _{i}\frac{\sqrt{2}}{3}
\right)  \label{P_a*} \\
(P_{0},P_{i}) &=&(Mc,P_{i})=-\frac{c^{2}}{2\sqrt{2}G}\left( \chi ,\chi _{i}%
\frac{\sqrt{2}}{3}\right) .  \nonumber
\end{eqnarray}

We find that: 
\begin{eqnarray}
\chi &=&-2\sqrt{2}\frac{G}{c^{2}}M  \label{chi} \\
\chi _{i} &=&6\frac{G}{c^{3}}P^{i}.
\end{eqnarray}

\section{Appendix C}

Relations between the following quantities have been used through this work:

\begin{eqnarray}
l^{a} &=&\frac{\sqrt{2}}{2}(1,\frac{\zeta +\overline{\zeta }}{1+\zeta 
\overline{\zeta }},-i\frac{\zeta -\overline{\zeta }}{1+\zeta \overline{\zeta 
}},\frac{-1+\zeta \overline{\zeta }}{1+\zeta \overline{\zeta }}),
\label{tetrad} \\
m^{a} &=&\text{\dh }l^{a}=\frac{\sqrt{2}}{2}(0,\frac{1-\overline{\zeta }^{2}%
}{1+\zeta \overline{\zeta }},\frac{-i(1+\overline{\zeta }^{2})}{1+\zeta 
\overline{\zeta }},\frac{2\overline{\zeta }}{1+\zeta \overline{\zeta }}), 
\nonumber \\
\overline{m}^{a} &=&\overline{\text{\dh }}l^{a}=\frac{\sqrt{2}}{2}(0,\frac{%
1-\zeta ^{2}}{1+\zeta \overline{\zeta }},\frac{i(1+\zeta ^{2})}{1+\zeta 
\overline{\zeta }},\frac{2\zeta }{1+\zeta \overline{\zeta }}),  \nonumber \\
t^{a} &=&\sqrt{2}(1,0,0,0), \\
n^{a} &=&t^{a}-l^{a}=\frac{\sqrt{2}}{2}(1,-\frac{\zeta +\overline{\zeta }}{%
1+\zeta \overline{\zeta }},i\frac{\zeta -\overline{\zeta }}{1+\zeta 
\overline{\zeta }},\frac{1-\zeta \overline{\zeta }}{1+\zeta \overline{\zeta }%
}),  \nonumber \\
c^{a} &=&l^{a}-n^{a}=\sqrt{2}(0,\frac{\zeta +\overline{\zeta }}{1+\zeta 
\overline{\zeta }},-i\frac{\zeta -\overline{\zeta }}{1+\zeta \overline{\zeta 
}},\frac{-1+\zeta \overline{\zeta }}{1+\zeta \overline{\zeta }}).  \nonumber
\end{eqnarray}
Letting a,b, etc., take the values $(0,1,2,3)=(0,i)$ we have the products

\begin{eqnarray*}
(m_{i}\overline{m}_{j}-\overline{m}_{i}m_{j}) &=&i\frac{\sqrt{2}}{2}\epsilon
_{ijk}c_{k}, \\
(m_{i}c_{j}-m_{j}c_{i}) &=&-i\sqrt{2}\epsilon _{ijk}m_{k}, \\
m_{i}c_{j} &=&\frac{1}{2}(m_{i}c_{j}-m_{j}c_{i})+\frac{1}{2}%
(m_{i}c_{j}+m_{j}c_{i}) \\
&=&-i\frac{1}{\sqrt{2}}\epsilon _{ijk}m_{k}+(l=2)Terms
\end{eqnarray*}

In a recent article\cite{TsH} we have developed a notation to describe
symmetric and trace-free tensor products of the three Euclidean vectors (c$%
_{i}$,$m_{i},\overline{m}_{i}$) and their products;

Several examples that have been used in the text are: 
\begin{eqnarray*}
c_{i} &=&-Y_{(1)i}^{(0)} \\
m_{i} &=&Y_{(1)i}^{(1)} \\
\overline{m}_{i} &=&Y_{(1)i}^{(-1)} \\
m_{i}m_{j} &=&Y_{(2)ij}^{(2)} \\
c_{i}m_{j}+m_{i}c_{j} &=&-Y_{(2)ij}^{(1)} \\
3c_{i}c_{j}-2\delta _{ij} &=&Y_{(2)ij}^{(0)}
\end{eqnarray*}
where the upper index indicates the spin weight and the lower index in
parenthesis gives the $l$ value. The indices i and i,j, etc., are Euclidean
tensor indices.

Some examples of product decompositions (Clebsch-Gordon expansion) that have
been used in the text are: 
\begin{eqnarray*}
Y_{1i}^{1}Y_{1j}^{0} &=&\frac{i}{\sqrt{2}}\epsilon _{ijk}Y_{1k}^{1}+\frac{1}{
2}Y_{2ij}^{1}, \\
Y_{1k}^{1}Y_{1f}^{-1} &=&\frac{1}{3}\delta _{kf}-\frac{i}{2\sqrt{2}}\epsilon
_{kfl}Y_{1l}^{0}-\frac{1}{12}Y_{2kf}^{0}.
\end{eqnarray*}

We have found that this type of notation and analysis is extremely useful.

\end{document}